\documentclass[prb, twocolumn, superscriptaddress]{revtex4-2}
\usepackage{bm, amsmath, amsfonts, amssymb, mathtools, ascmac, braket}
\usepackage{multirow}
\usepackage{graphicx}
\usepackage{float, xcolor}
\usepackage{physics}
\usepackage{comment}

\usepackage[
pagebackref=false,
colorlinks=true,
linkcolor=blue,
urlcolor=blue,
filecolor=black,
citecolor=red,
pdfstartview=FitV,
pdftitle={},
pdfauthor={},
pdfsubject={},
pdfkeywords={},
pdfpagemode=None,
bookmarksopen=true
]{hyperref}

\newcommand{\ii}{\text{i}}

\begin{document}

\title{Multifractal statistics of non-Hermitian skin effect on the Cayley tree}

\author{Shu Hamanaka}
\email{hamanaka.shu.45p@st.kyoto-u.ac.jp}
\affiliation{Department of Physics, Kyoto University, Kyoto 606-8502, Japan}
\affiliation{Institute for Theoretical Physics, ETH Zurich, 8093 Zurich, Switzerland}
\affiliation{Department of Physics, University of Zurich, 8057 Zurich, Switzerland}

\author{Askar A. Iliasov}
\affiliation{Department of Physics, University of Zurich, 8057 Zurich, Switzerland}

\author{Titus Neupert}
\affiliation{Department of Physics, University of Zurich, 8057 Zurich, Switzerland}

\author{Tom\'a\v{s} Bzdu\v{s}ek\ }
\affiliation{Department of Physics, University of Zurich, 8057 Zurich, Switzerland}

\author{Tsuneya Yoshida}
\affiliation{Department of Physics, Kyoto University, Kyoto 606-8502, Japan}
\affiliation{Institute for Theoretical Physics, ETH Zurich, 8093 Zurich, Switzerland}
\affiliation{Department of Physics, University of Zurich, 8057 Zurich, Switzerland}

\date{\today}

\begin{abstract}
Multifractal analysis is a powerful tool for characterizing the localization properties of wave functions. 
Despite its utility, this tool has been predominantly applied to disordered Hermitian systems. 
Multifractal statistics associated with the non-Hermitian skin effect remain largely unexplored. 
Here, we demonstrate that the tree geometry induces multifractal statistics for the single-particle skin states on the Cayley tree by deriving the analytical expression of multifractal dimensions.
This sharply contrasts with the absence of multifractal properties for conventional single-particle skin effects in crystalline lattices.
Our work uncovers the unique feature of the skin effect on the Cayley tree and provides a novel mechanism for inducing multifractality in open quantum systems without disorder.
\end{abstract}

\maketitle

\section{Introduction}
Multifractal analysis is an effective tool for describing the localization properties of wave functions commonly used in disordered Hermitian systems. 
A prime example is Anderson localization, where single-particle wave functions are localized by disorder~\cite{Anderson58}.
In three or higher dimensions, wave functions exhibit multifractal behavior at critical points~\cite{Evers-RMP,Wegner1980,Castellani_1986,Schreiber-PRL-1991,Mudry-96,Evers-PRL-2000}.
High-dimensional limits can be effectively described by the Bethe lattice (infinite Cayley tree), the simplest regular tree graph. 
The Anderson transition on the Bethe lattice~\cite{rabou-73,*rabou-74} has played a pivotal role in multifractal analysis as its loop-free structure enables analytical derivations of critical exponents of multifractal wave functions~\cite{Efetov-85,*Efetov-87,Zirnbauer-86-1, Mirmin-91}.
Recently, multifractal analysis of Anderson localization on graphs has garnered renewed attention~\cite{Tikonov-review-21,Monthus-09, *Monthus-11,Luca-14,Tikonov-PRB-16,Mata-PRL-17,Baroni-PRB-24} due to its connection with many-body localization~\cite{Nandkinshore-15,Abanin-RMP-19,Alet-18}, since the Fock space has a locally tree-like structure~\cite{Altshuler-PRL-97}.

A different mechanism of localization is offered by the non-Hermitian skin effect~\cite{Yao-18, Kunst-18, Okuma-23}.
This phenomenon originates from nonreciprocal dissipation, 
where a macroscopic number of bulk states are localized.
In one dimension, most of the eigenstates are exponentially localized,
while in higher dimensions, skin effects are diversified~\cite{Hofmann-20, Zhang-22} including the higher-order skin effect~\cite{Kawabata-PRB-20,Okugawa-20,Li-23,Fu-21,ZhangXiu-2021}.
The non-Hermitian skin effect has recently gained attraction because of its relation to non-Hermitian topology~\cite{Gong-18, Kawabata-19, Zhou-19, Bergholtz-21, Yokomizo-PRL-2019, Kawabata-symple-20, Amoeba-24, Wojcik2020, Okuma-20, Zhang-20,Borgnia-20,Zirnstein-21,Kawabata-21,Yoshida-20, Longhi-19,Longhi-22, Nakamura-24, Schindler-23, Nakamura-23,Ma-24, hamanaka-nonhinh-24, nakamura-wannier-24, HN-many-2-22, HNmany-1-22, Faugno-22, Okuma-prl-21,Tanaka-24, Sun-21, Nakai-24, Wang-Yi-23, Manna-23,Guo-PRB-23,Kim-24,nakagawa-arxiv-24} and significant influence on open quantum dynamics~\cite{Song-19,Liu-20,Haga-21,*marko-15,Mori-20,kawabata-23,Yang-22,Feng-24,hamanaka-23,Jiang-24,Begg-PRL-24,shen-23,Yoshida-24,Jacopo-24,Ekman-24,shimomura-24, Kuo-24,Soumya-24,Pietro-24,Yoshida-PRA-23,marche-arxiv-24}.
Experimental observations have been reported in both open classical and quantum systems~\cite{Brandenbourger-19,Ghatak-20,ZhangXiu-2021,Helbig-20,Hofmann-20,Xiao-20,Liang-PRL-2022,Gou-PRL-2020}.

Despite extensive studies on the non-Hermitian skin effect, the application of multifractal analysis remains largely unexplored~\cite{hamanaka-2024}.
In particular, multifractal statistics of single-particle non-Hermitian skin effects have not been addressed.
Conventionally, single-particle skin effects in crystalline lattices lack multifractal properties since skin modes always occupy a finite fraction of the system trivially. 
However, following the spirit by viewing many-body localization as Anderson localization on the hierarchical lattice~\cite{Altshuler-PRL-97,Tikonov-review-21}, the 
emergence of multifractality in the many-body skin effect~\cite{hamanaka-2024} suggests that the single-particle skin effect on a tree-like graph exhibits multifractality.

In this work, we demonstrate that the single-particle skin effect displays multifractal statistics on the Cayley tree.
We derive the analytical expression of multifractal dimensions and elucidate that multifractality emerges as a direct consequence of the tree geometry and skin effect.
Specifically, the exponential growth of the Hilbert space dimensions with respect to the layer number and exponential localization of skin modes leads to intricate scaling behavior.
This property starkly contrasts with the absence of multifractality
for conventional skin effects on crystalline lattices. Our
work thus reveals the unique feature of the skin effect on
the tree graph.

The paper is organized as follows.
In Sec.~\ref{sec: multifractal}, using the moments of the wave function, we introduce multifractal statistics, which play a central role in this paper.
Section~\ref{sec: model} describes the model that we analyze in this study.
Specifically, we study the nonreciprocal Hamiltonian on the Cayley tree with connectivity $K$.
In Sec.~\ref{sec:K2}, we demonstrate that a subset of eigenstates on the Cayley tree, which we dub symmetric eigenstates, display multifractal statistics for $K \geq 2$.
We also show that the symmetric eigenstates are special when compared to the remaining (non-symmetric) eigenstates in that their multifractal dimensions are robust against weak disorder.
In Sec.~\ref{sec: discuss}, we conclude our work with several outlooks.
In Appendix~\ref{app:hn}, we review the eigenvalue equation of the Hatano-Nelson model~\cite{Hatano-PRL-1996, *Hatano-PRB-1997}, which is related to the Cayley tree with the trivial connectivity $K=1$.
In Appendix~\ref{app:K1}, we solve the eigenvalue equation exactly for $K=1$ and demonstrate that these eigenstates are characterized by zero multifractal dimensions.
In Appendix~\ref{app:K2}, we provide the complete solutions of the eigenvalue equations for the Cayley tree for an arbitrary $K \geq 2$ and show the detailed calculation of multifractal dimensions. 
In Appendix~\ref{app:higher-order}, we demonstrate the absence of multifractality of single-particle skin effects in conventional crystalline lattices.
In Appendix~\ref{app:strong}, we discuss effects of strong disorder on multifractal dimensions.
\section{Multifractal analysis}
\label{sec: multifractal}
In this section, we summarize 
the basic notions characterizing the multifractal scaling of the wave function (see also Sec. II C of Ref.~\cite{Evers-RMP}).
We consider an $\mathcal{N}$-components normalized wave function 
$\ket{\psi}$ in a given basis $\{ \ket{j} \}~{(j=1,\cdots,\mathcal{N})}$, $\ket{\psi} = \sum_{j=1}^\mathcal{N} \psi_j \ket{j}$. 
Multifractality of the wave function is characterized by an infinite set of exponents of its moments.
Specifically, the moment $I_q$ (inverse participation ratio) defined by~\cite{Wegner1980}
\begin{align}
    I_q \coloneqq \sum_{j=1}^{\mathcal{N}} \abs{\psi_j}^{2q}
\end{align}
follows the scaling $I_q \propto \mathcal{N}^{-\tau_q}$ with a non-decreasing ($\tau_q^\prime \geq 0$) and convex ($\tau_q^{\prime\prime} \leq 0$) function $\tau_q$ satisfying $\tau_0 = -1$ and $\tau_1 = 0$.
Multifractal dimensions $D_q$ defined from the exponents $\tau_q$
\begin{align}
    \label{eq: Dq}
    D_q \coloneqq \frac{\tau_q}{q-1} 
\end{align}
quantify the effective dimensions of the wave function occupying the Hilbert space~\footnote{
Following Refs.~\cite{Luca-14,Mace-19}, we define $\tau_q$ from the scaling of $I_q \propto \mathcal{N}^{-\tau_q}$  with respect to the dimension of the Hilbert space $\mathcal{N}$. This notation differs from Ref.~\cite{Evers-RMP}, where $\tau_q$ is defined through the scaling $I_q \propto L^{-\tau_q}$ for a $d$-dimensional system with the side length 
$L$. In the latter notation, $D_q$ takes values from $0$ to $d$.
We also note $D_1$ is defined by the differential coefficient of $\tau_q$ at $q=1$.}.
We focus on $q \geq 0$ where multifractal dimensions satisfy $0 \leq D_{q} \leq 1$.
For perfectly delocalized states, we have $D_q = 1$.
In contrast, when the states are localized in a finite region of the Hilbert space, we have $D_q = 0$.
In the intermediate regime $0 < D_q < 1$, the wave function is extended but not delocalized.
The state is fractal when $D_q$ takes a constant value in the intermediate regime.
If $D_q$ depends on $q$, the wave function possesses multifractal statistics,
forming an intricate distribution within Hilbert space.
This complexity necessitates a continuous (multiple) set of exponents to fully characterize the scaling of its moments.
The multifractal spectrum $f_\alpha$ defined via the Legendre transformation 
\begin{align}
\label{eq:fa}
    f_\alpha \!\coloneqq \alpha q  \!-\! \tau_q ~~{\rm at~} q {~\rm s.t.~~}\!  \lim_{\epsilon\to 0^+}\!\!\left. \frac{d \tau}{dq}\right|_{q+\epsilon} \!\!\!\!\!\leq \!\alpha\! \leq\!\!  \lim_{\epsilon\to 0^+} \!\! \left. \frac{d \tau}{dq}\right|_{q-\epsilon}\!\!\!
\end{align}
describes how different scaling parameters $\alpha$ are distributed in the multifractal system~\cite{Halsey-PRA-86}.
For delocalized states, $f_1 = 1$ and $f_{\alpha \neq 1} = -\infty$ otherwise, forming a needle shape in the $(\alpha,f_\alpha)$-plane.
In contrast, the broadened needle and finite support $ \alpha_{\rm min} < \alpha < \alpha_{\rm max}$ where $f_\alpha > 0$ gives a signature of multifractality.

Prime examples where wave functions show different multifractal statistics are provided by disordered Hermitian systems.
The Anderson localized phase is characterized by $D_q = 0$, while the delocalized states follow $D_q = 1$.
Multifractality $0 < D_q < 1$ appears for the Anderson transition~\cite{Evers-RMP,Wegner1980,Castellani_1986,Schreiber-PRL-1991,Mudry-96,Evers-PRL-2000} and the many-body localized phase~\cite{Luitz-15,Kravtsov-15,Serbyn-17,Mace-19,Monteiro-PRR-21,De-PRB-21}. 
Another example of the multifractal behavior is given by the many-body non-Hermitian skin effect~\cite{hamanaka-2024}, whereas the single-particle skin effect is perfectly localized in one dimension ($D_q = 0$).

\section{Model}
\label{sec: model}
\begin{figure}[t]
\centering
\includegraphics[width=0.85\linewidth]{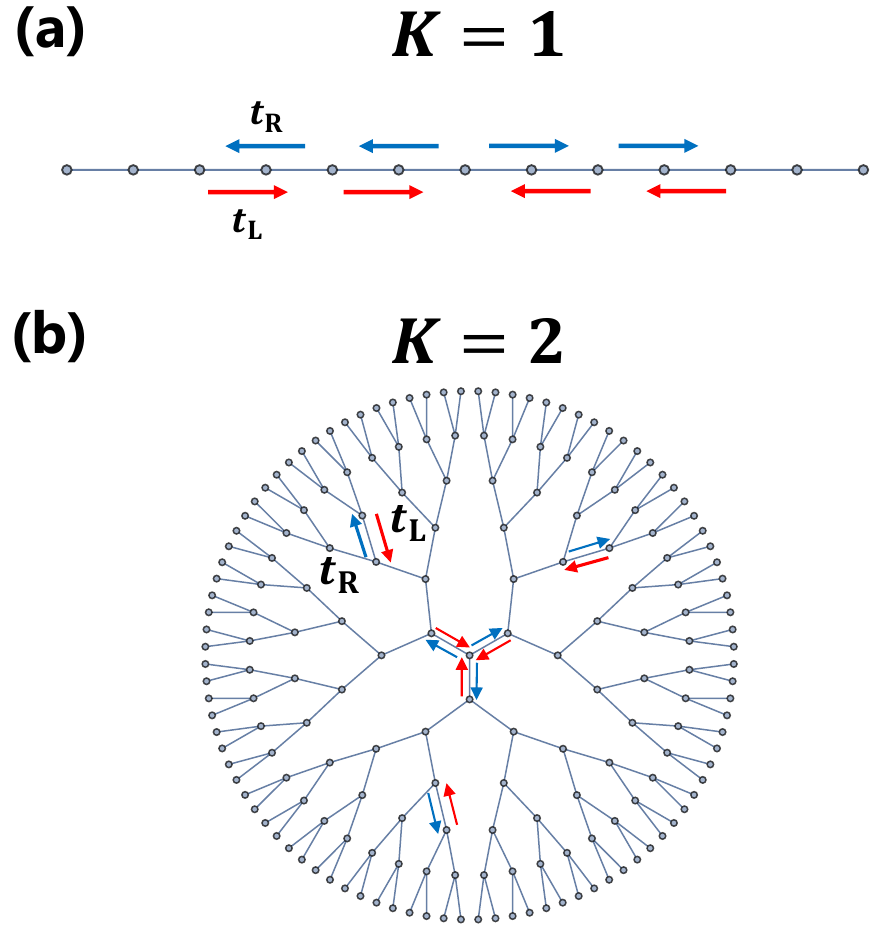} 
\caption{
A sketch of the model with nonreciprocal hopping on the Cayley tree for (a) $K=1$ and (b) $K=2$. The number of layers is $M=6$. The hopping amplitude from the boundary to the center (from the center to the boundary) is given by~$t_{\rm L}$~($t_{\rm R}$).
}
\label{fig:graph}
\end{figure}
In order to study the non-Hermitian skin effect on tree graphs, we consider the nonreciprocal Hamiltonian defined on the Cayley tree (Fig.~\ref{fig:graph}).
The Cayley tree is a tree graph whose nodes have the same branch number $K$  (i.e., each site has $K+1$ neighbors) except for the surface nodes.
We generate the Cayley tree as follows. 
First, we define the central node.
Then we generate the first layer consisting of $K+1$ nodes and connect them with the central node.
Subsequently, we create the next layer by attaching $K$ distinct nodes to each ``parent'' node in the previous layer (Fig.~\ref{fig:graph}).
Repeating the final procedure $M$ times, we obtain the $M$-layer Cayley tree with connectivity $K$.
Each layer consists of a set of nodes that are equidistant from the central node.
The Cayley trees with $M=6$ layers for $K=1$ and $K=2$ are given in Fig.~\ref{fig:graph} (a) and (b) respectively.

To realize the non-Hermitian skin effect, we consider the nonreciprocal Hamiltonian on the Cayley tree with connectivity $K$ described by
\begin{align}
\label{eq:hamiltonian}
    H =  t_{\rm R} \sum_{\langle i > j \rangle}
      \ket{i} \bra{j} 
    + t_{\rm L}
    \sum_{\langle i < j \rangle}    \ket{i} \bra{j}
\end{align}
with hopping amplitudes $t_{\rm R}, t_{\rm L} > 0 $. 
Nonreciprocity arises when $\beta \coloneqq \sqrt{t_{\rm R}/t_{\rm L}}$ differs from one ($t_{\rm R} \neq t_{\rm L}$).
The symbol $\sum_{\langle i > j \rangle}$ \big($\sum_{\langle i < j \rangle}$\big) represents the summation over neighboring nodes $i$ and $j$, where the node $j$ is closer to (farther from) the central node than the node $i$ (see Fig.~\ref{fig:graph}).
For $t_{\rm R} < t_{\rm L} ~(t_{\rm R} > t_{\rm L})$, nonreciprocal hopping tends to carry the particle toward the central node (surface nodes).
While skin effects in non-crystalline lattices have been recently explored~\cite{Manna-23,Wang-Yi-23,Soumya-24,Shi-PRB-24}, the previous approaches cannot capture intricate scaling behavior of skin modes.
This is because $q$-dependent multifractal dimensions are necessary to distinguish the skin mode in non-crystalline lattices from that in crystalline lattices.
Our works present the first demonstration of multifractal statistics of the single-particle skin effect.

Experimental realizations of non-Hermitian tight-binding models can be found in both quantum and classical systems. 
In quantum systems, non-Hermitian Hamiltonians describe dynamics between quantum jump events in the quantum trajectory method~\cite{Daley-Ad-2014, Ashida-adv-20, Nakagawa-PRL-20, Honda-PRL-23}.
Recently, the non-Hermitian skin effect induced by asymmetric hopping has been reported in ultracold atoms~\cite{Liang-PRL-2022,Gou-PRL-2020}.
In classical systems, non-Hermiticity can arise due to resistors in electric circuits or general losses and friction in mechanical or acoustic systems. The former can be controlled to a large degree, which led to the realization of various non-Hermitian phenomena in electric circuits including the skin effect~\cite{Helbig-20,Hofmann-20}. In addition to losses, one can also include gain through active elements~\cite{Kotwal-21}.

\begin{figure}[t]
\centering
\includegraphics[width=\linewidth]{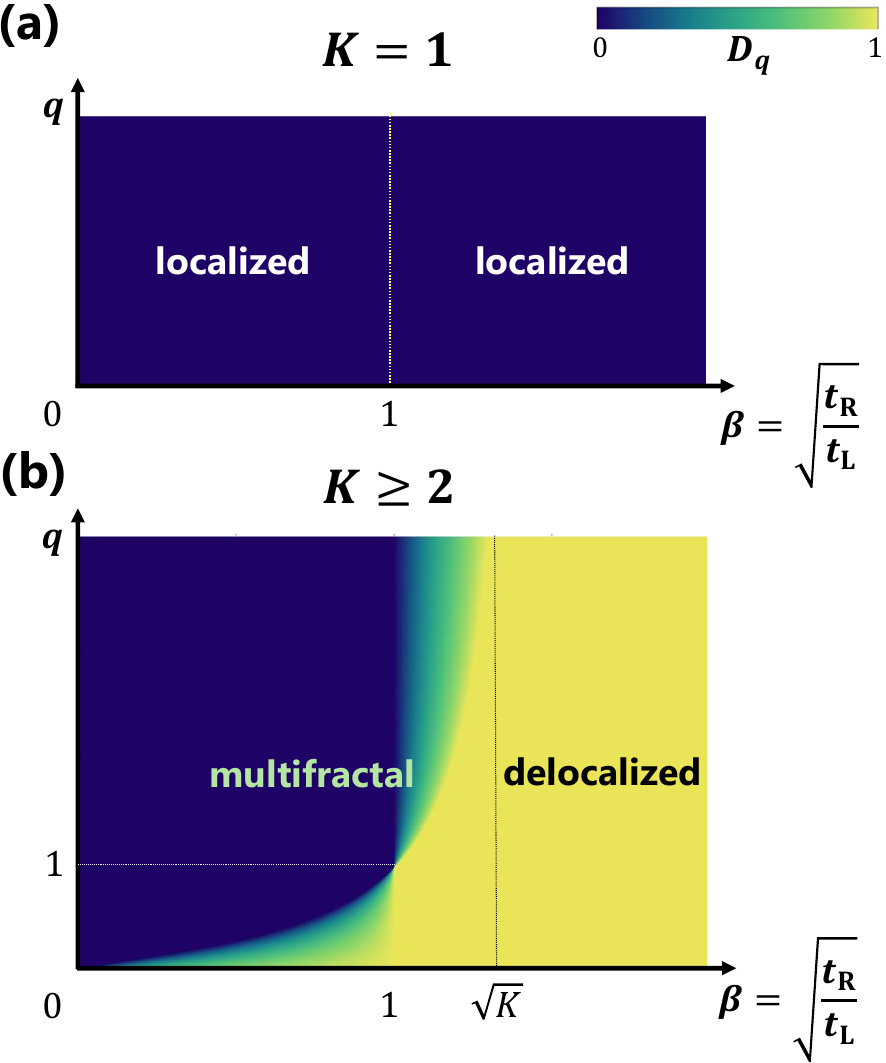}
\caption{
Phase diagrams of multifractal dimensions $D_q$. 
(a) For $K=1$, the eigenstates exhibit localization $D_q = 0$ for any nonreciprocity $\beta \neq 1$. (b) For $K \geq 2$, while the symmetric eigenstates are localized in the limit $\beta \rightarrow 0$, they are delocalized for $\beta > \sqrt{K}$. In the intermediate regime $0<\beta< \sqrt{K}$, multifractality appears.
}
\label{fig:phase}
\end{figure}

\section{Results}
\label{sec:K2}
Before turning to our main case of interest $K \geq 2$, let us briefly outline the result for the special case $K=1$.
For $K=1$, the Cayley tree is just a one-dimensional chain. The eigenstates of the Hamiltonian in Eq.~\eqref{eq:hamiltonian} are exponentially localized in the presence of any non-Hermiticity $\beta \neq 1$. (Note that the directionality of the hopping changes at the central node, so the $K=1$ Cayley tree does not represent a finite segment of a uniform chain.)
Since such localized states exhibit $D_q = 0$ (see Appendix~\ref{app:K1}), the eigenstates do not possess the multifractal property [Fig.~\ref{fig:phase} (a)].

We now demonstrate multifractal statistics for the Cayley tree with connectivity $K \geq 2$.
In Sec.~\ref{subsec:K2-basis}, we generate the appropriate basis states so that the eigenvalue equation is analytically solved.
Then, we derive eigenvalues and eigenstates of the non-Hermitian Cayley tree in Sec.~\ref{subsec:K2-recur}, focusing in particular on a class of solutions that we refer to as symmetric eigenstates~\cite{Mahan-2001-PRB, Aryal-IOP-2020}.
Using these eigenstates, we compute multifractal dimensions in Sec.~\ref{subsec:K2-Dq}. Finally, in Sec. \ref{sec:weak-disorder}, we numerically study how the multifractal dimension (for concreteness we focus on $D_2$) of the eigenstates is affected by their energy degeneracy and disorder.
We find that while the multifractal dimension of the symmetric eigenstates is uniquely determined and robust against weak disorder, the remaining states (referred to as non-symmetric states) exhibit extensive energy degeneracies that make their multifractal dimensions sensitive to disorder.

\begin{figure*}[t]
\centering
\includegraphics[width=0.93\linewidth]{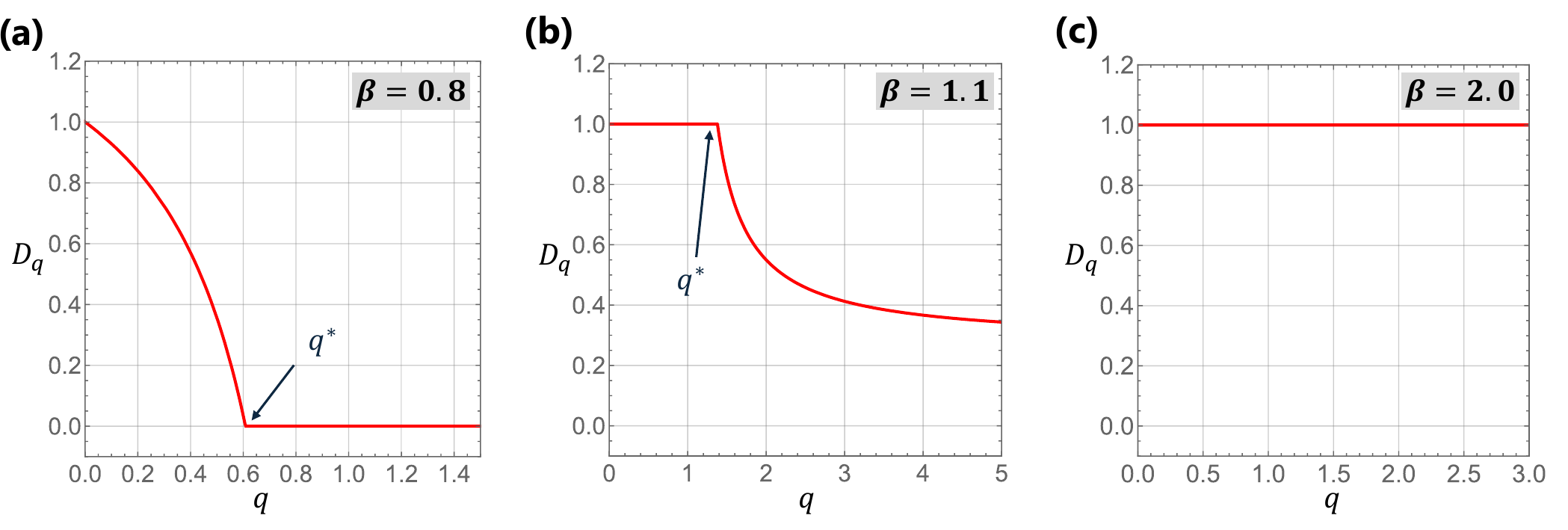} 
\caption{The dependence of the multifractal dimensions $D_q$ on $q$ for (a) $0 < \beta 
<1 $, (b) $1 < \beta 
< \sqrt{K}$, (c) $\sqrt{K} < \beta 
$.
The data are
plotted for the case $K=2$, and $\beta$ is chosen as (a) $\beta=0.8$, (b) $\beta=1.1$, (c) $\beta=2.0$.
}
\label{fig:Dq}
\end{figure*}

\subsection{Symmetric basis states}\label{subsec:K2-basis}
We denote the total number of layers as $M$.
Since the central node has $K+1$ branches and the $l$-th $(l=1, \cdots, M)$ layer of each branch has $K^{l-1}$ nodes, the dimension $\mathcal N$ of the Hilbert space is 
\begin{align}
    \mathcal{N} =  1 + (K+1) \times \sum_{l=1}^M K^{l-1} = 1 + (K+1) \frac{K^M -1}{K-1}.
\end{align}
We construct the $(K+1)M+1$ symmetric basis states, which span a subspace of the whole Hilbert space as follows.  

First, we choose the central node as one of the symmetric basis states as
\begin{align}
    |0) \coloneqq \ket{0}.
\end{align}
In this paper, $\ket{\cdots}$ denotes the position basis and $|\cdots)$ denotes the symmetric (and the complementary non-symmetric) basis states. 

Second, we generate the remaining symmetric basis states $|l)_m$ by symmetrizing the position basis $\ket{l,j,m} (j=1, \cdots, K^{l-1})$ in the $l$-th $(l=1,\cdots,M)$ layer of the branch $m$ ($m= 1,\cdots,K+1$) as
\begin{align}
  | l)_m \coloneqq \frac{1}{\sqrt{K^{l-1}}} \sum_{j=1}^{K^{l-1}}  \ket{l,j,m},
\end{align}
which form $(K+1)M$ orthonormal states [see Fig.~\ref{fig:basis}~(a) in Appendix~\ref{app:K2}].
We call eigenstates that can be decomposed using only the symmetric basis states as symmetric eigenstates, i.e., they are expanded as
\begin{align}\label{eq:symmetric}
    \ket{\Psi} = \psi_0 |0) + \sum_{l=1}^M \sum_{m=1}^{K+1} \psi_{l,m} |{l})_m,
\end{align}
where $\psi_0$ and $\psi_{l,m}$ are the wave function components.
We show in Appendix~\ref{app:anyK-basis} how $|0)$ and $| l)_m$ can be supplemented by further basis states, which we call non-symmetric basis states, so that their union forms a complete orthonormal basis of the Hilbert space with dimension $\mathcal{N}$.

\subsection{Symmetric eigenstates}\label{subsec:K2-recur}
In the subspace spanned by the symmetric basis states, the eigenvalue equation $H \ket{\Psi} = E \ket{\Psi}$ is reduced to recurrence relations.
The eigenvalue equation has $KM$ linearly independent 
solutions with $\psi_{0} = 0$ (the other $M+1$ solutions with $\psi_0 \neq 0$ are given in Appendix~\ref{app:K2}), determined~by 
\begin{align}
\begin{split}
\label{eq:ev-sym-psi0}
    E \psi_{l,m} &= \sqrt{K} t_{\rm R} \psi_{l-1,m} +  \sqrt{K} t_{\rm L} \psi_{l+1,m} \\
     \psi_{0,m} &= \psi_{M+1,m} = 0 \\
    0 &= \sum_{m=1}^{K+1} \psi_{1,m}
    \end{split}
\end{align}
with $\psi_{0,m} \coloneqq \psi_0/ \sqrt{K}$~\footnote{We introduce $\psi_{0,m}$ to achieve brevity of Eq.~\eqref{eq:ev-sym-psi0}.
} for $l=1,2,\cdots, M$ and $m=1,\cdots,K+1$.
The first and second lines in Eq.~\eqref{eq:ev-sym-psi0} are nothing but the eigenvalue equation of the Hatano-Nelson model~\cite{Hatano-PRL-1996, *Hatano-PRB-1997}, which is the prototypical model exhibiting the non-Hermitian skin effect.
As shown in Appendix~\ref{app:hn}, the solutions of the eigenvalue equation of the Hatano-Nelson model are given by
\begin{align}
\label{eq:sol-psi0}
    \psi_{l,m}^{(n)} = \beta^l \sin{(\theta_n l)}, \quad E_n = 2 \sqrt{K t_{\rm R} t_{\rm L}} \cos{\theta_n}
\end{align}
with $\beta = \sqrt{t_{\rm R} / t_{\rm L}}$ and $\theta_n = n \pi /(M+1) ~(n=1,2,\cdots,M)$.
Thus, the eigenstates take the form 
\begin{align}
\label{eq:sol-sym}
    \ket{\Psi_n} =  \sum_{m=1}^{K+1} \sum_{l=1}^M 
     c_m \psi_{l,m}^{(n)} |{l})_m
\end{align}
with constant values $c_{m} \in \mathbb{C} ~(m=1,\cdots,K+1)$.
Since the eigenstates in Eq.~\eqref{eq:sol-psi0}
do not depend on the branch number $m$, we have $K+1$ degenerate solutions for each $n$.  
From the third line of Eq.~\eqref{eq:ev-sym-psi0},
the values $c_m$ must satisfy $\sum_{m=1}^{K+1} c_m = 0$.
Therefore the number of degeneracy for each $n$ reduces to $K$, and thus there are $KM$ linearly independent solutions.
It should be noted that the obtained symmetric eigenstates are not degenerate with any of the remaining eigenstates if $M+1$ is chosen to be a prime number (see also Appendix~\ref{app:K2}).

Although every eigenstate in Eq.~(\ref{eq:sol-sym}) is $K$-fold degenerate, this degeneracy is independent of the layer number $M$. It follows that the multifractal dimensions, computed in next section~\ref{subsec:K2-Dq}, do not depend on the choice of the values $c_m$ and hence can be uniquely determined [see discussion around Eq.~\eqref{aeq:Dq01}].
As we will see in Sec.~\ref{sec:weak-disorder}, the symmetric eigenstates differ in this respect significantly from the non-symmetric eigenstates, which exhibit an extensive energy degeneracy.

\subsection{Multifractality}\label{subsec:K2-Dq}
Using the symmetric eigenstates in Eq.~\eqref{eq:sol-sym}, 
we obtain the inverse participation ratio $I_q$ in the position basis~$\ket{l, j, m}$
\begin{align}
\label{eq:Iq-sym}
    I_q=  \left[
    \frac{\beta^2 -1}{ \beta^2 (\beta^{2M}-1)}
    \right]^q
    \frac{
    1-\left(\frac{\beta^{2q}}{K^{q-1}}\right)^M 
    }{\beta^{-2q}-{K^{1-q}}} \sum_{m=1}^{K+1} \abs{c_m}^{2q}.
\end{align}
Importantly, while the dimension of the Hilbert space increases exponentially as $\mathcal{N} \propto K^M$ due to the inherent structure of the Cayley tree, 
the inverse participation ratio can also increase exponentially but with different base values, either as $I_q \propto \beta^{-2qM}$ or $I_q \propto  {(\beta^{2q}/K^{q-1})}^M$ due to the skin effect.
This indicates that the skin modes on the tree intricately occupy the Hilbert space with various scale structures. 
Below, we quantitatively characterize the complexity of these skin modes, arising from the interplay between the geometry of the tree and the skin effect, by providing analytical expressions of the $q$-dependent multifractal dimensions.
Depending on $\beta$, the multifractal dimensions take different forms 
in the following three cases (see Appendix~\ref{app:K2}).

For $0 < \beta < 1$, we have 
 \begin{equation}\label{eq:Dq1}
       D_q = 
       \begin{cases}
           1 - \frac{q}{q-1} \frac{\log (\beta^2)}{\log K} \quad &(q <q^*) \\
           0 \quad &(q > q^*)
       \end{cases}
\end{equation}
with the dimensionless parameter 
\begin{align}
    q^* \coloneqq \frac{\log{K}}{\log{K} - \log{(\beta^2)}}.
\end{align}
Since the dimensions $D_q$ depends on $q$, multifractality appears in this regime~[Fig.~\ref{fig:Dq} (a)]. 
In the limit $\beta \rightarrow 0$, $q^*$ approaches zero ($q^* \rightarrow 0$), leading to $D_{q} = 0$ for all $q$. 
The vanishing of $D_{q}$ indicates localization of eigenstates around the central node, induced by strong nonreciprocity ($ t_{\rm L} \gg t_{\rm R}$).
Recall that an infinitesimal non-Hermiticity can induce localization for the one-dimensional chain ($K=1$) [Fig.~\ref{fig:phase} (a)].
In contrast, the existence of localized states for $K \geq 2$ requires limit $t_{\rm R}/ t_{\rm L} \rightarrow 0$ [Fig.~\ref{fig:phase} (b)], which stems from the structural difference of the Cayley tree between $K=1$ and $K \geq 2$.

For $1 < \beta < \sqrt{K}$, we have
 \begin{equation}\label{eq:Dq2}
       D_q = 
       \begin{cases}
           1 \quad &(q <q^*) \\
          \frac{q}{q-1} \frac{\log (\beta^2)}{\log K} \quad &(q > q^*).
       \end{cases}
   \end{equation}
As in the previous regime, multifractality appears~[Fig.~\ref{fig:Dq}~(b)].
In the limit $\beta \rightarrow \sqrt{K}$, $q^*$ diverges ($q^* \rightarrow \infty$), leading to $D_q = 1$ for all $q$.
This describes the perfect delocalization of eigenstates throughout the Cayley tree.
In contrast to localization ($D_q=0$) requiring the limit $\beta \rightarrow 0$, the delocalized state emerges for finite $\beta$. 
The transition point $\beta = \sqrt{K}$ coincides with the square root of connectivity. 
Physically, this means that the particle cannot be pushed to the graph surface until the outward hopping amplitude $t_{\rm R}$ exceeds $K$ times the inward hopping amplitude $t_{\rm L}$ due to the branching of the incoming wave into 
$K$ parts at each node.

For $\beta > \sqrt{K}$, we have
\begin{equation}\label{eq:Dq3}
       D_q = 1
\end{equation}
which indicates that the eigenstates are delocalized~[Fig.~\ref{fig:Dq} (c)].
In the limit $\beta \rightarrow\infty$, the eigenstates spread over the surface of the tree. 
Although eigenstates are accumulated at individual disconnected nodes on the tree, most of the nodes reside at the surface in the thermodynamic limit.
Since the multifractal dimension quantifies the effective dimension of the wave function occupancy in the Hilbert space, these states are characterized by $D_q=1$.


\begin{figure}[t]
\centering
\includegraphics[width=\linewidth]{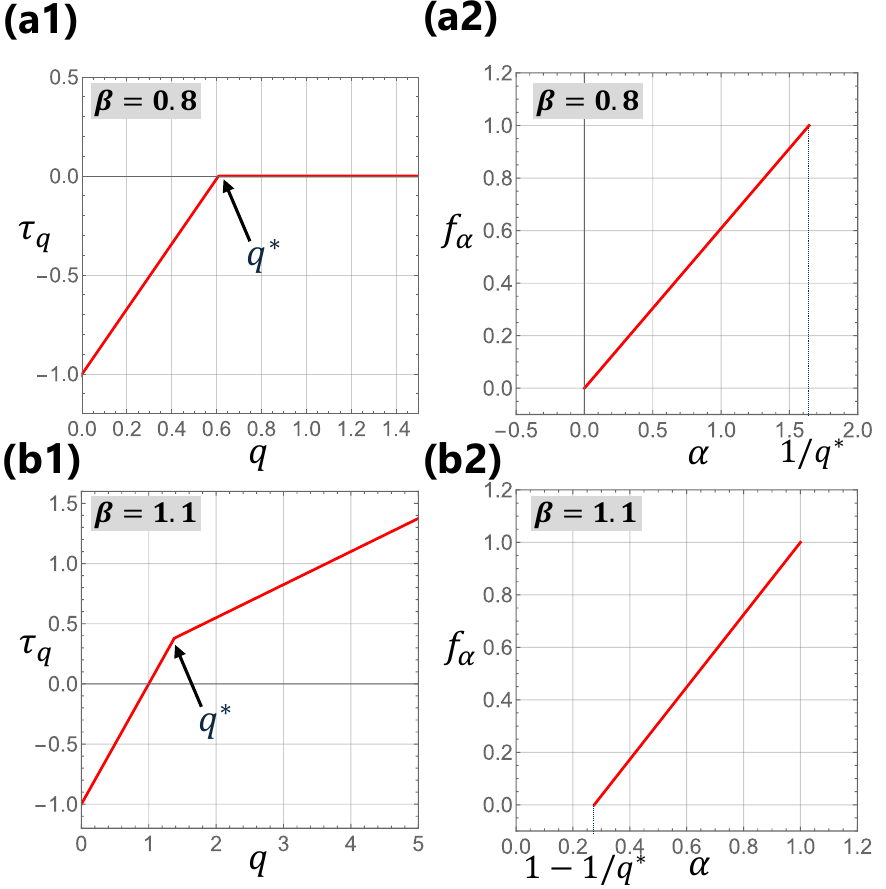} 
\caption{The dependence on $q$ of the exponent
$\tau_q$, and the dependence on $\alpha$
of the multifractal spectrum $f_\alpha$, displayed for $\beta = 0.8$ (a1, a2) and for $\beta = 1.1$ (b1, b2).
The data are
plotted for the case $K=2$.
}
\label{fig:fq}
\end{figure}
For both $0 < \beta < 1 $ and $ 1 < \beta < \sqrt{K}$, symmetric eigenstates display multifractal statistics.
However, their $q$-dependence is crucially different [Fig.~\ref{fig:Dq} (a,b)].
To illustrate this difference, we further calculate the exponents $\tau_q$ and the multifractal spectrum $f_\alpha$ obtained as (see Appendix~\ref{app:K2})
\begin{align}
\label{eq:a-tauq}
    \tau_q = 
    \begin{cases}
        \frac{1}{q^*}(q-q^*) ~&(q <q^*) \\
        0 ~&(q > q^*),
    \end{cases} 
\end{align}
\begin{align}
     f_\alpha = 
    \begin{cases}
        q^* \alpha ~&( 0 < \alpha < 1 / q^*) \\
        - \infty ~&(\rm{ otherwise})
    \end{cases}
\end{align}
for $0 < \beta <1$, and
\begin{align}
  \label{eq:b-tauq}
      \tau_q = 
      \begin{cases}
         q-1  ~&(q <q^*) \\
           \frac{\log (\beta^2)}{\log K} q ~&(q > q^*),
      \end{cases} 
\end{align}
\begin{align}
f_\alpha = 
      \begin{cases}
          q^* \alpha + (1-q^*) ~ &\left( 1-1/q^* < \alpha < 1 \right) \\
          - \infty ~ &(\rm{ otherwise})
      \end{cases}
\end{align}
for $1 < \beta < \sqrt{K}$.

Importantly, for $ 0 < \beta < 1$, $\tau_q$ becomes zero when $q > q^*$ holds [Fig.~\ref{fig:fq} (a1)]. For the Anderson transition in $d$ dimension, a sparse character of the wave function localized around particular nodes leads to $\tau_q = 0$ when $q>1/2$ in the limit $d \rightarrow \infty$~\cite{Mildenberger-02, EFETOV1990119,* Fyodorov1992, Mirlin94a, *Mirlin94b}.
The exponent $\tau_q$ in Eq.~\eqref{eq:a-tauq} implies that the skin modes possess similar multifractal statistics, although the threshold depends on $q^*$.
Given a strongly localized wave function $\ket{\psi}$, its moment $I_q$ for large $q$ is dominated by the largest components $\abs{\psi_j}$, leading to $\tau_q = 0$. 
This observation implies that the states are concentrated around the central node for $0<\beta<1$, consistent with Eq.~\eqref{eq:sol-psi0}.
On the other hand, for small $q$, which quantifies the average degree of localization, the state exhibits multifractal statistics.
The multifractal spectrum $f_\alpha$ takes supremum at $\alpha \simeq 1/q^*$ [Fig.~\ref{fig:fq}~(a2)], which indicates that the wave functions locally scale as $\abs{\psi_j}^2 \propto \mathcal{N}^{-1/q^*}$ in many places. 
Note in passing that the abrupt change of the derivative of $\tau_q$ at $q=q^*$  
[or equivalently, the triangular shape of the multifractal spectrum $f(\alpha)$] 
implies the phase transition at this point.
Indeed, the vanishing of the exponent $\tau_q = 0$ for finite $q$ is known as {\it freezing transition}.
While such a frozen phase was observed in the Anderson transition on Bethe lattice~\cite{Luca-14}, the two-dimensional random Dirac model~\cite{Chamon-PRL-96}, and the Rosenzweig-Porter random matrix model~\cite{Kravtsov-15}, but in contrast, we reveal this unique multifractality arising from the skin effect, which is attributed to nonreciprocity rather than disorder.

In contrast, for $ 1 < \beta < \sqrt{K}$, $\tau_{q}$ is always positive when $q>1$ holds [Fig.~\ref{fig:fq}~(b1)].
This is contrary to $\tau_q=0$ with large $q$ for Anderson localization in infinite dimensions and for the previous $0<\beta<1$ case, implying unique multifractality of the non-Hermitian skin effect on the Cayley tree.
The appearance of multifractality at large $q$ reflects the absence of strongly localized peaks in the wave function.
Since the multifractal spectrum $f_\alpha$ takes supremum at $\alpha \simeq 1$ [Fig.~\ref{fig:fq}~(b2)], the wave functions scale as $\abs{\psi_j}^2 \propto \mathcal{N}^{-1}$  almost everywhere.
This is consistent with $D_q=1$ for small $q$ in Fig.~\ref{fig:Dq}~(b).

It should be noted that the conventional single-particle skin effects in crystalline lattices do not possess multifractal properties.
Let us consider a state which occupies an 
$n$-dimensional area in a $d$-dimensional lattice (e.g., $(d-n)$-th order non-Hermitian skin mode~\cite{Kawabata-PRB-20,Okugawa-20}). 
Then the multifractal dimensions are estimated as $D_q \simeq n/d$ (see Appendix~\ref{app:higher-order}). While the fractality $0<D_q <1$ can appear, this state is characterized by a single exponent and free from ``multifractality''.
In contrast, the skin modes on the Cayley tree indeed exhibit $q$-dependent multifractal dimensions, suggesting a unique feature of the skin effect on the tree graph.

\begin{figure*}[t]
\centering
\includegraphics[width=1\linewidth]{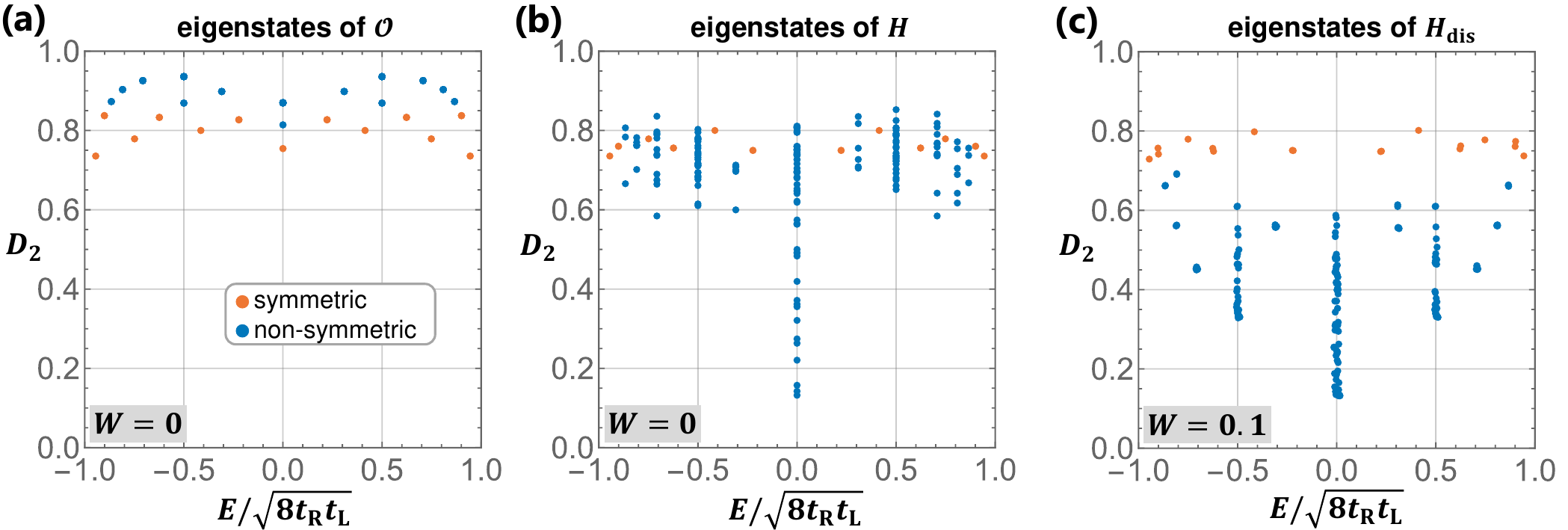} 
\caption{The multifractal dimension $D_2$ as a function of the rescaled energy $E/\sqrt{8 t_{\rm R}t_{\rm L}}$ for a Cayley tree with $K=2$, $M+1=7$, $t_{\rm R}=1.1$ and $t_{\rm L}=0.9$ ($\beta=\sqrt{11/9}$) computed: (a,b) in the absence of disorder ($W=0$), resp.~(c) in the presence of weak disorder ($W=0.1$). 
Data for symmetric (non-symmetric) eigenstates are shown in orange (blue).
In panel (a), simultaneous eigenstates of the commuting set $\mathcal{O}$ [Eq.~\eqref{eqn:operator-set}] are considered, which are homogeneously distributed over all branches of the Cayley tree.
In panel (b), we consider generic eigenstates of the Hamiltonian $H$ without constructing specific linear combinations.
}
\label{fig:disorder}
\end{figure*}

Multifractality of the Anderson transition is usually accompanied by the divergence of the localization length~\cite{Evers-RMP}. 
In contrast, the skin modes on the Cayley tree are exponentially localized with the localization length $1 / \xi =  \log{(\sqrt{K}/\beta)}$ in the entire multifractal phase $0 < \beta < \sqrt{K}$.
This is also opposed to the power law localized skin modes of the symplectic Hatano-Nelson model near criticality~\cite{kawabata-23} and the critical skin modes characterized by the system size dependent localization length~\cite{Li-20,Yokomizo-21}.
Furthermore, for the Anderson transition on a regular lattice, multifractal wave functions exhibit intricate spatial distributions. 
In contrast, the wave functions in Eq.~\eqref{eq:sol-psi0} spread monotonically from the central node to the surface. 
Hence, multifractal statistics originate from the tree geometry rather than from a complex distribution of the
wave functions.
Therefore our results reveal the unique feature of the skin effect on the tree graph and the novel mechanism of inducing multifractality in open systems.

\subsection{
Multifractal dimensions of degenerate states
}\label{sec:weak-disorder}

Up to now, we have focused on the small set of symmetric eigenstates specified by Eq.~\eqref{eq:sol-psi0} and on their multifractal properties, because the remaining eigenstates on the Cayley tree exhibit extensive energy degeneracy.
We refer to these additional eigenstates as non-symmetric eigenstates because they cannot be expanded using the symmetric basis states according to Eq.~\eqref{eq:symmetric}.
By forming linear combinations of these extensively degenerate non-symmetric eigenstates, it is possible to obtain a new set of orthogonal eigenstates that are characterized by a potentially different value of the moments $I_q$. 
We find that, as a consequence, the multifractal dimensions of the non-symmetric eigenstates are not well defined.


In this section, we first explicitly demonstrate the dependence of multifractal dimensions  $D_{q=2} = -\log I_{q=2} /\log \mathcal{N}$ in finite size systems on the choice of the degenerate eigenstates, taking $K=2$ as an example.
We construct eigenstates through two approaches: one using a symmetry-adapted method, where linear combinations are chosen so that eigenstates are homogeneously distributed over all branches of the Cayley tree, and the other without taking specific combinations.
We observe that the former approach yields a larger $D_2$ compared to the latter.
Second, we consider the effects of weak disorder, which are expected to arise in real experimental systems.
We show that weak disorder favors more localized linear combinations of the degenerate eigenstates, leading to a lower value of $D_2$, while leaving $D_2$ of nondegenerate symmetric eigenstates unaffected.

First, we demonstrate that the multifractal dimension $D_2$ depends on the choice of degenerate eigenstates.
Specifically, we show that a symmetry-adapted construction yields a larger $D_2$ compared to a generic construction without taking specific linear combinations. 
To construct linear combinations of the degenerate eigenstates that are distributed over all branches of the system, we rely on the symmetry of the Cayley tree with a coordination number equal to three [Fig.~\ref{fig:graph} (b)].
We consider the following symmetry operators:
One is the rotation $C_3$, which cyclically permutes the three branches that split off the central site $\ket{0}$.
Next, we consider the swap $C_{2,l}$ (defined for each $l\in\{1,\ldots,M-1\}$), which corresponds to the simultaneous exchange of the pairs of branches that split off all sites in $l$-th layer. 
Repeated application of these operators brings the sites to their initial arrangement, namely the cubic power of the rotation $(C_3)^3=\mathbf{1}$ and the square of the swaps $(C_{2,l})^2=\mathbf{1}$ yield the identity.
Since these $M$ symmetries preserve the shape of the Cayley tree, they commute with the Hamiltonian~(\ref{eq:hamiltonian}): 
\begin{equation}
[C_3,H]=0=[C_{2,l},H].
\end{equation}
In addition, as the symmetries permute the sites of the system at various depths, they commute with each other: 
\begin{equation}
[C_3,C_{2,l}]=0=[C_{2,l},C_{2,l'}].
\end{equation}
Therefore, the collection 
\begin{equation}
\label{eqn:operator-set}
\mathcal{O}=\{H, C_3, C_{2,1},\ldots C_{2,M-1}\}
\end{equation}
forms a commuting set of operators, and they can be diagonalized simultaneously.
In other words, it is possible to construct eigenstates of $H$ that also possess well-defined eigenvalues for each of the listed symmetries.

Since a suitable composition of the symmetries $C_3$ and $C_{2,l}$ can translate a given site to any other site in the same layer, non-symmetric eigenstates of $H$ that are simultaneously eigenstates of all the operators in $\mathcal{O}$ have a uniform probability distribution within the individual layers.
This implies enhanced delocalization of such simultaneous eigenstates over generic non-symmetric eigenstates of $H$, for which the computer arbitrarily selects a linear combination of degenerate states, thereby failing to enforce homogeneity within the layers. Note that the Hamiltonian commutes with operators in $\mathcal{O}$ for both cases.

To test this prediction, we plot the multifractal dimension $D_2$ of the numerically obtained simultaneous eigenstates of $\mathcal{O}$ in Fig.~\ref{fig:disorder} (a), where orange (blue) corresponds to the symmetric (non-symmetric) eigenstates.
For comparison, Fig.~\ref{fig:disorder} (b) displays $D_2$ for generic eigenstates of the Hamiltonian $H$ (without constructing symmetry-adapted linear combinations).
The difference between the two panels indicates that $D_2$ of degenerate states depends on their construction and thus on the choice of the linear combination of degenerate states.
Furthermore, the non-symmetric eigenstates of $\mathcal{O}$ display a large value of $D_2$ compared to generic eigenstates of $H$.
Specifically, we observe a drop of blue dots from $D_2\in[0.8,0.95]$ in the symmetry-adapted method to $D_2 \in [0.1,0.9]$ in the generic method.
We find the same value of $D_2$ for nondegenerate symmetric eigenstates in both Fig.~\ref{fig:disorder} (a,b).
Symmetric eigenstates with $\psi_0=0$, which are two-fold degenerate, exhibit a relatively small change of their multifractal dimension within the range $D_2 \in [0.75,0.85]$.
This small deviation converges for taking a sufficiently large $M$ since $D_q$ do not depend on the choice of degenerate states [see discussion around Eq.~\eqref{aeq:Dq01}].
We note that although non-symmetric eigenstates yield the same multifractal dimension when specific linear combinations are chosen (see Appendix \ref{app:K2}), Fig.~\ref{fig:disorder} (b) demonstrates that multifractality arises for generic eigenstates. This implies that multifractality is a general property of the skin modes on the tree lattice, irrespective of the choice of linear combinations.

Second, we demonstrate that the multifractal dimension of symmetric eigenstates is robust to weak disorder, in contrast to the sensitivity observed in highly degenerate non-symmetric eigenstates. 
We consider the Hamiltonian with random on-site potentials:
\begin{align}
    H_\textrm{dis}=H + \sum_{j=1}^\mathcal{N} U_j \ket{j}\bra{j},
\end{align}
where $H$ is the nonreciprocal Hamiltonian in Eq.~(\ref{eq:hamiltonian}) and $U_j$ takes a random value for each site ($U_j \in [-u, u]$, $u=W\Delta$ with $W \geq 0$ and the energy interval $\Delta = \sqrt{4K t_{\rm R}t_{\rm L}}/M$).
Figure~\ref{fig:disorder} (c) shows the numerically computed values of $D_2$ for weak disorder $W=0.1$.
Since disorder breaks the rotation and swap symmetries of the Cayley tree, the symmetry-adapted computation is no longer applicable, and we should compare against Fig.~\ref{fig:disorder}~(b) where we adopted generic eigenstates of the Hamiltonian without disorder ($W=0$).
The comparison of the data in Fig.~\ref{fig:disorder} (b, c) confirms that while $D_2$ of symmetric eigenstates exhibits only negligible variation to weak disorder, $D_2$ of the highly degenerate non-symmetric states drops significantly.
Due to the general tendency of disorder to drive (Anderson) localization, the weak disorder leads to such linear combinations of the degenerate non-symmetric eigenstates that are concentrated on particular branches of the Cayley tree.
In contrast, since such linear combinations are not available for the nondegenerate symmetric eigenstates, their $D_2$ is robust against weak disorder.

Note that the above discussion can be generalized to any $K\geq 2$.
First, for the symmetry considerations, the larger values of $K$ necessitate a replacement of the three-fold rotation $C_3$ of the branches splitting off site $\ket{0}$ by a $(K+1)$-fold rotation $C_{K+1}$, and the replacement of the two-fold swaps $C_{2,l}$ of the sub-branches at layer $l$ by $K$-fold cyclic permutations $C_{K,l}$. 
These adjustments furnish an adapted set of commuting operators, whose simultaneous eigenstates are uniformly distributed within the individual layers of the Cayley tree with connectivity $K$, producing non-symmetric eigenstates with an enhanced value of $D_2$.
Second, when considering the role of weak disorder, our argument remains unchanged.
On the one hand, since the disorder strength cannot exceed the energy interval of the symmetric eigenstates, their multifractal dimensions are robust against weak disorder.
On the other hand, as the degeneracy of the non-symmetric eigenstates is enhanced with larger $K$, we expect their multifractal dimensions to exhibit an even stronger sensitivity to weak disorder.

\section{Discussions}
\label{sec: discuss}
In this paper, we have demonstrated that the tree geometry induces multifractal statistics for the single-particle non-Hermitian skin effect (Fig.~\ref{fig:phase}).
Specifically, the symmetric eigenstates display multifractal statistics on average ($q \simeq 1$) for $0<\beta<1$.
For $1<\beta<\sqrt{K}$, the states locally scale as $\abs{\psi_j}^2 \propto \mathcal{N}^{-1}$ almost everywhere, showing multifractal statistics only in high-moment regimes.
In the presence of strong nonreciprocity $\beta>\sqrt{K}$, the eigenstates are delocalized, originating from the expander property of the Cayley tree.
We have also shown the multifractal dimension of these symmetric eigenstates remains robust to weak disorder as opposed to the sensitivity observed in highly degenerate non-symmetric eigenstates.
Since conventional exponentially localized single-particle skin modes in crystalline lattices do not possess multifractal properties, we have provided a novel mechanism for inducing multifractality. 

While we have focused on the simplest graph in this paper, our findings could serve as a benchmark for understanding the many-body non-Hermitian skin effect~\cite{hamanaka-2024}, as analytically determining the eigenstates of such many-body systems is difficult. 
In this respect, multifractal analysis of the skin effect on loop graphs, including random graphs~\cite{Tikonov-review-21,Tikonov-PRB-16} is important, as they can more accurately capture the intricate structure of the Fock space.
Furthermore, although multifractality appears off criticality in this study, it is noteworthy that the symplectic Hatano-Nelson model exhibits a nonequilibrium phase transition~\cite{kawabata-23}. 
The skin effect near criticality should display different multifractal statistics.

More broadly, since multifractality on the non-Hermitian Cayley tree can be traced to the exponential growth of the system with radius, it should be interesting to generalize our model to other lattices sharing this property, known as expander graphs~\cite{Laumann:2009}.
As a noteworthy example, regular hyperbolic lattices~\cite{Boettcher:2022} are expander graphs which were in recent years realized in several experimental platforms~\cite{Kollar:2019,Lenggenhager:2022, Huang:2024}.
Finally, and more speculatively, we point out the feature of Fig.~\ref{fig:disorder} (c), where a small number of isolated multifractal eigenstates appear within a sea of localized eigenstates. 
This behavior is reminiscent of scar states~\cite{Sanjay-PRB-18}, appearing in an interacting quantum system, which are isolated states with sub-volume-law entanglement, embedded within a dense spectrum of volume-law-entangled excited states.
Motivated by this apparent similarity, we anticipate novel insights could be gained by examining the entanglement entropy of multifractal states in our and related setups.

\medskip
{\it Note added.\,---\,}After completion of this work, we became aware of a recent related work~\cite{sun-arxiv-24} that investigates the Hatano-Nelson model on an iterative lattice. We also note another recent work~\cite{Hatano-arxiv-2024} which focuses on transport properties on a tree lattice featuring complex potentials.

\medskip
\section*{acknowledgments}
We thank Manfred Sigrist for discussions.
S.H. thanks Hideaki Obuse, Kohei Kawabata and Shoki Sugimoto for discussions.
We thank Masatoshi Sato and Kenji Shimomura for kindly informing us that they independently considered the non-Hermitian network model from a different perspective after our initial submission~\cite{private}.
S.H. is supported by JSPS Research Fellow No.~24KJ1445 and JSPS Overseas Challenge Program for Young Researchers.
S.H. and T.Y are grateful for the support and hospitality of the Pauli Center for Theoretical Studies.
A.I. acknowledges support from the UZH Postdoc Grant, grant No. FK-24-104.
A.I. and T.B. were supported by the Starting Grant No. 211310 by the Swiss National Science Foundation.
T.N. acknowledges support from the Swiss National Science Foundation through a Consolidator Grant (iTQC, TMCG-2 213805).
T.Y is supported by the Grant from Yamada Science Foundation.
This work is supported by JSPS KAKENHI Grant Nos.~JP21K13850 and JP23KK0247, JSPS Bilateral Program No.~JPJSBP120249925.

\appendix


\section{Hatano-Nelson model}
\label{app:hn}
We review the solutions of the eigenvalue equation for the Hatano-Nelson model~\cite{Hatano-PRL-1996,*Hatano-PRB-1997}, which is the typical model exhibiting the non-Hermitian skin effect (see also  Sec. SI of the Supplemental Material in Ref.~\cite{Yokomizo-PRL-2019} and Sec. II in Ref.~\cite{Chen-PRB-23}).
We consider the Hatano-Nelson model under open boundary conditions:
\begin{align}
\label{aeq:K1-hn-hamiltonian}
    H_{\rm HN} = \sum_{j=1}^{L-1} \Big( t_{\rm R} \ket{j+1}\bra{j} + t_{\rm L} \ket{j}\bra{j+1} \Big).
\end{align}
The eigenvalue equation $H_{\rm HN} \ket{\psi} = E \ket{\psi}$ is reduced to 
\begin{align}
    & t_{\rm R} \psi_{j-1} + t_{\rm L} \psi_{j+1} = E \psi_{j} \quad (j=1,\cdots,L) \label{aeq:K1-hn-bulk}\\
    & \psi_0 =\psi_{L+1} = 0. \label{aeq:K1-hn-edge}
\end{align}
Using the ansatz $\psi_j \propto \beta^j$, Eq.~\eqref{aeq:K1-hn-bulk} becomes
\begin{align}
    t_{\rm L} \beta^2 -E \beta + t_{\rm R} = 0,
\end{align}
which has two solutions $\beta = \beta_\pm$ satisfying 
\begin{align}
\label{aeq:K1-tasu}
    \beta_+ +\beta_- = \frac{E}{t_{\rm L}}, \quad \beta_+ \beta_- = \frac{t_{\rm R}}{t_{\rm L}}.
\end{align}
From Eq.~\eqref{aeq:K1-hn-edge}, the general solution $\psi_j = c_+ \beta_+^j + c_- \beta_-^j$ satisfies 
\begin{align}
    c_{+} + c_{-} = 0 \\
     c_{+}\beta_+^{L+1} + c_{-}\beta_-^{L+1} = 0
\end{align}
leading to 
\begin{align}
    0 = c_{+} (\beta_+^{L+1} - \beta_-^{L+1}),
\end{align}
and hence
\begin{align}
    \frac{\beta_+}{\beta_-} = e^{2 \ii \theta_n}, \quad \theta_n =  \frac{n \pi}{L+1} \quad (n=1,2,\cdots,L).
\end{align}
We rewrite $\beta_\pm$ as
\begin{align}
    \beta_\pm = \beta e^{\pm \ii \theta_n}.
\end{align}
From the second equation in Eq.~\eqref{aeq:K1-tasu}, we obtain $\beta = \sqrt{t_{\rm R}/t_{\rm L}}$, leading to
\begin{align}
    \psi_j \propto \beta^j (e^{\ii\theta_n j} - e^{-\ii\theta_n j}) \propto \beta^j \sin{(\theta_n j)}.
\end{align}
From the first equation Eq.~\eqref{aeq:K1-tasu}, the eigenvalues become
\begin{align}
    E_n = 2 \sqrt{t_{\rm R} t_{\rm L}} \cos{\theta_n}.
\end{align}
Thus, in the presence of non-Hermiticity $\beta \neq 1$,
all eigenstates are localized around the boundary.

\section{Absence of multifractality for $K=1$}
\label{app:K1}
\subsection{Eigenstates}\label{app:K1:eigenstates}
\begin{figure}[t!]
\centering
\includegraphics[width=\linewidth]{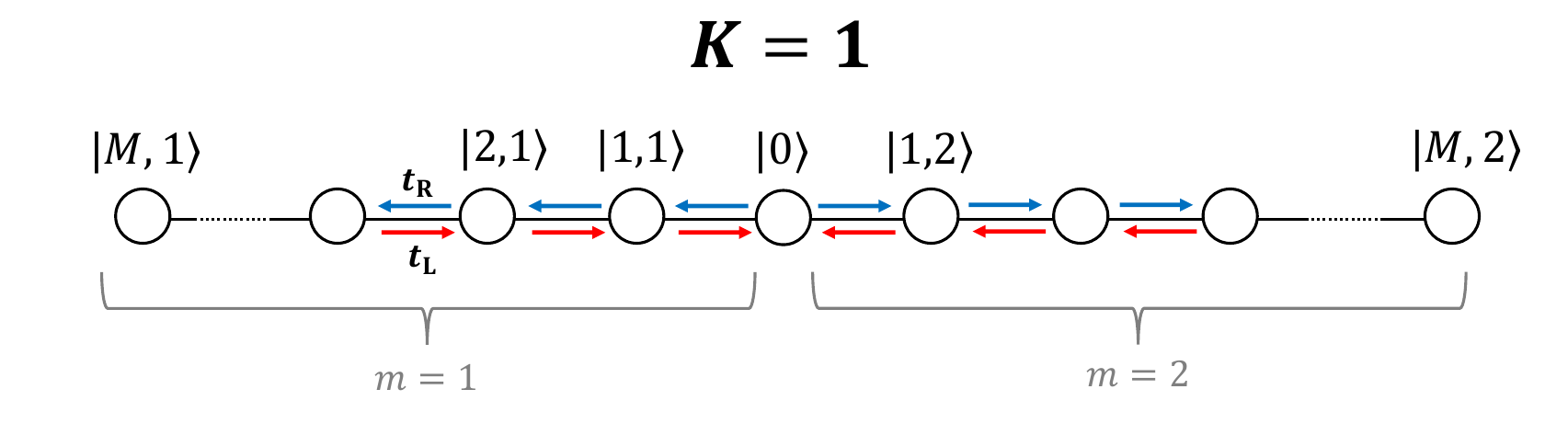} 
\caption{The Cayley tree with connectivity $K=1$. The hopping from the center to the outer layer (from the outer layer to the center) is $t_{\rm R}$ ($t_{\rm L}$).}
\label{afig:1d}
\end{figure}

We demonstrate that eigenstates of the single-particle Hamiltonian on the Cayley tree with connectivity $K=1$ exhibit the perfect localization ($D_q=0$).
The eigenvalue equation $H \ket{\psi} = E \ket{\psi}$ with
\begin{align}
\label{aeq:hamil-K1}
    H \!=\! \sum_{j=0}^{M-1} \! \sum_{m=1}^2 \!\! \Big( \! t_{\rm R} \! \ket{j\!+\!1, \! m}\!\bra{j, m} \!+\! t_{\rm L} \!\ket{j, m}\!\bra{j\!+\!1, m} \! \Big)
\end{align}
is reduced to
\begin{align}
\begin{split}
\label{aeq:hn-bulk}
     E \psi_{j,m} &= t_{\rm R} \psi_{j-1,m} + t_{\rm L} \psi_{j+1,m}  \\
     E \psi_{0} &= t_{\rm L} (\psi_{1,1} + \psi_{1,2} ) \\
     0 & = \psi_{M+1,m}
\end{split}
\end{align}
for $j=1,\cdots,M$ and $m=1,2$, where we have expanded the state as~\footnote{The central node is shared, so wave function at origin is identified as $\psi_{0} \coloneqq \psi_{0,1} = \psi_{0,2} $.} 
\begin{align}
    \ket{\psi} = \psi_{0} \ket{0} + \sum_{j=1}^M \sum_{m=1}^2 \psi_{j,m} \ket{j,m}.
\end{align}

Since the Hamiltonian commutes with the inversion operator $\mathcal{I}$ defined by 
\begin{align}
    \mathcal{I} = \ket{0}\bra{0} + \sum_{j=1}^M \Big[ \ket{j,1}\bra{j,2} +  \ket{j,2}\bra{j,1} \Big]
\end{align}
with respect to the central node (Fig.~\ref{afig:1d}), we can take their simultaneous eigenstates. 
Given that $\mathcal{I}^2=1$, the eigenvalues of the inversion operator are $I=\pm 1$.
In the following, we determine the eigenstates for the two cases where the inversion operator has eigenvalues $I=-1$ and $I=+1$.

\begin{enumerate}
    \item \textit{The case of $I=-1$} \\\\     
Since the eigenvalue of the inversion operator is $I=-1$,
we have 
\begin{align}
    \psi_0 =0,~ \psi_{j,1} = -\psi_{j,2}.
\end{align}
Then Eq.~\eqref{aeq:hn-bulk} reduces 
to
\begin{align}
\begin{split}
\label{aeq:hn-psi0}
     E \psi_{j,m} &= t_{\rm R} \psi_{j-1,m} + t_{\rm L} \psi_{j+1,m}  \\
     0 &  = \psi_{0,m} = \psi_{M+1,m} 
\end{split}
\end{align}
for $j=1,\cdots,M$.
Using the procedure described in Appendix~\ref{app:hn}, the solutions of the first and second lines of Eq.~\eqref{aeq:hn-psi0} are 
\begin{align}
\label{aeq:sol-psi0}
    \psi_{j,m}^{(n)} \!=\! \beta^j \sin{(\theta_n j)}, \;\; E_n \!=\! 2 \sqrt{t_{\rm R} t_{\rm L}} \cos{\theta_n}
\end{align}
with $ \theta_n = n \pi / (M+1)  ~(n=1,2,\cdots,M)$ and $\beta = \sqrt{t_{\rm R} / t_{\rm L}}$.
Therefore, we obtain $M$ linearly independent solutions
\begin{align}
\label{aeq:K1-sol-antisym}
    \ket*{\psi^{(n)}} =  \sum_{j=1}^M \Big[ \psi_{j,1}^{(n)} \ket{j,1} - \psi_{j,2}^{(n)} \ket{j,2} \Big].
\end{align}

\item \textit{The case of $I=+1$}  \\\\
Since the eigenvalue of the inversion operator is $I=+1$,
we have 
\begin{align}
     \psi_{j,1} = \psi_{j,2}.
\end{align}
Then Eq.~\eqref{aeq:hn-bulk} reduces
to
\begin{align}
\begin{split}
\label{aeq:hn-psi-neq0}
    E \psi_{j,m} &=  t_{\rm R} \psi_{j-1,m} + t_{\rm L} \psi_{j+1,m} \\
    E \psi_{0} &= 2 t_{\rm L} \psi_{1,m}  \\
    0 &= \psi_{M+1,m}
    \end{split}
\end{align}
for $j=1,2,\cdots, M$.
We use the ansatz
\begin{align}
\label{aeq:ansatz}
    E &= 2 \sqrt{t_{\rm L} t_{\rm R}} \cos{\theta}, \nonumber \\
    \psi_0 &= \sin{\gamma}, \\
    \psi_{j,m} &= \beta^j \sin{( \theta j + \gamma)}  \nonumber
\end{align}
with unknown parameters $\theta,\gamma$.
The ansatz in Eq.~\eqref{aeq:ansatz} satisfies the first line of Eq.~\eqref{aeq:hn-psi-neq0}.
From the second line, we obtain
\begin{align}
\label{aeq:K1-gamma-omega1}
    \sin{\theta} \cos{\gamma} = 0.
\end{align}
From the third line of Eq.~\eqref{aeq:hn-psi-neq0}, we have
\begin{align}
\label{aeq:K1-gamma-omega2}
    \sin{\left[ \theta (M+1)  + \gamma \right]} = 0.
\end{align}
Combining Eqs.~\eqref{aeq:K1-gamma-omega1} and Eq.~\eqref{aeq:K1-gamma-omega2}, we obtain $M+1$ solutions for $\theta_n$ and $\gamma_n ~(n=1,\cdots,M+1)$, given by
\begin{align}
    \quad \theta_n = \Big(n-\frac{1}{2}\Big) \frac{\pi}{M+1}, \quad \gamma_n = \frac{\pi}{2}.
\end{align}
Thus, the solutions are given by
\begin{align}
    \psi_{j,m}^{(n)} = \beta^j \cos{(\theta_n j)} \quad (j=0,1,\cdots,M).
\end{align}
Therefore, we obtain $M+1$ linearly independent solutions
\begin{align}
\label{aeq:ansatz-sol}
     \ket*{\psi^{(n)}} =  \psi_0 \ket{0} +  
     \sum_{j=1}^M \Big[ \psi_{j,1}^{(n)} \ket{j,1} + \psi_{j,2}^{(n)} \ket{j,2} \Big]. \nonumber \\
\end{align}
\end{enumerate}

\subsection{Perfect localization ($D_q = 0$)}
\label{app-sub: Dq=0 K=1}
We now compute the multifractal dimensions $D_q$, defined by 
\begin{align}
    D_q = \lim_{\mathcal{N}
    \to\infty}\frac{1}{1-q} \frac{1}{\log{\mathcal{N}}} \log{I_q},
\end{align}
from the inverse participation ratio
\begin{align}
    I_q =  \sum_{j,m} \abs{ \bra*{{j,m}}\ket*{\psi}}^{2q},
\end{align}
where $\mathcal{N} = 2 M +1$ is the Hilbert space dimension (i.e., the system size).
Our analysis demonstrates that the eigenstates obtained in Appendix~\ref{app:K1:eigenstates} exhibit zero fractal dimension ($D_q = 0$).
\begin{enumerate}
    \item\textit{Case of eigenstates~\eqref{aeq:K1-sol-antisym}} \\\\  
In Eq.~\eqref{aeq:sol-psi0}, we first approximate 
\begin{align}
\label{aeq:I-wv-approx-K1}
    \psi_{j,m}^{(n)} \simeq (\beta e^{\ii \theta_n})^j.
\end{align}
This approximation essentially follows the procedure of the non-Bloch band theory~\cite{Yao-18,Yokomizo-PRL-2019} and should accurately capture the nature of the skin effect (see also Appendix A in Ref.~\cite{hamanaka-2024}). 
This simplification leads to
\begin{align}
\ket*{\psi^{(n)}} \simeq
    \sum_{j=1}^M 
    \frac{1}{\sqrt{2 C}}
    \Bigg[
    (\beta e^{\ii \theta_n})^j 
    \Big(
    \ket{j,1} - \ket{j,2}
    \Big)
    \Bigg]
\end{align}
with the normalization constant
\begin{align}
    C =  \sum_{j=1}^M \beta^{2j} = \frac{ \beta^2 (\beta^{2M}-1)}{\beta^2 -1}.
\end{align}
Hence the inverse participation ratio $I_q$ becomes
\begin{align}
    I_q 
    &= \sum_{j,m} \abs{ \bra*{{j,m}}\ket*{\psi^{(n)}}}^{2q} \nonumber \\
    &=  \frac{2}{(2 C)^q} \sum_{j=1}^M \beta^{2qj}  \nonumber \\
    &= 2 \left[
    \frac{\beta^2 -1}{2 \beta^2 (\beta^{2M}-1)}
    \right]^q
    \frac{\beta^{2q}(\beta^{2qM}-1)}{\beta^{2q}-1}.\label{aeq:IPR-I}
\end{align}
Depending on the value $\beta$, we have two cases.
\begin{enumerate}
    \item[1)] For $\beta>1$, the factor 
    \begin{align}
        \frac{\beta^{2qM}-1}{(\beta^{2M}-1)^q}
    \end{align}
    in the inverse participation ratio $I_q$ is independent of $M$ in the limit $M \rightarrow \infty$. Thus we have $D_q = 0$.
    \item[2)] For $ \beta<1$, since $\beta^{2M} \rightarrow 0$
   in the limit $M \rightarrow \infty$, 
    the inverse participation ratio $I_q$ is independent of $M$.
    Thus we have $D_q  = 0$.
\end{enumerate}

\item \textit{Case of eigenstates~\eqref{aeq:ansatz-sol}} \\\\
In Eq.~\eqref{aeq:ansatz-sol}, we again approximate 
\begin{align}
    \psi_{j,m}^{(n)}  \simeq (\beta e^{\ii \theta_n})^j,
\end{align}
leading to
\begin{align}
    \ket*{\psi^{(n)}} \simeq
    \frac{1}{\sqrt{C}}
    \Bigg[
    \ket{0}
    +
    \sum_{j=1}^M 
    (\beta e^{\ii \theta_n})^j 
    \Big(
    \ket{j,1} + \ket{j,2}
    \Big)    
    \Bigg]
\end{align}
with the normalization constant
\begin{align}
    C = 1 + 2 \sum_{j=1}^M \beta^{2j} = \frac{\beta^2-1 + 2 \beta^2 (\beta^{2M}-1)}{\beta^2 -1}.
\end{align}
Hence, the inverse participation ratio $I_q$ becomes
\begin{align}
    I_q 
    &= \sum_{j,m} \abs{ \bra*{{j,m}}\ket*{\psi^{(n)}}}^{2q} \nonumber \\
    &=  \frac{2}{C^q} \Bigg[1+ \sum_{j=1}^M \beta^{2qj}  \Bigg] \nonumber \\
    &= 2 \left[
    \frac{\beta^2 -1}{ \beta^2-1 + 2 \beta^2 (\beta^{2M}-1)}
    \right]^q
     \Bigg[1 + 
    \frac{\beta^{2q}(\beta^{2qM}-1)}{\beta^{2q}-1}
    \Bigg]. \label{eq:IPR-I}
\end{align}
For both cases $\beta>1$ and $\beta<1$, similar calculations conducted below Eq.~\eqref{aeq:IPR-I} leads to
\begin{align} 
D_q = 0
\end{align}
in the limit $M \rightarrow \infty$.
\end{enumerate}
Therefore, all eigenstates of Hamiltonian~\eqref{aeq:hamil-K1} are perfectly localized (i.e., $D_q = 0$) in the single-particle Hilbert space.

\section{Complete solution for an arbitrary $K$}
\label{app:K2}
We obtain the complete set of eigenstates of the Hamiltonian in Eq.~\eqref{eq:hamiltonian} for an arbitrary $K \geq 2$ and compute their multifractal dimensions.
In Sec.~\ref{app:anyK-basis}, we provide the appropriate basis set by generalizing the discussion in Refs.~\cite{Mahan-2001-PRB,Aryal-IOP-2020} to solve the eigenvalue equation analytically.
Section~\ref{app:anyK-recrel} provides the recurrence relations for the chosen basis states.
In Sec.~\ref{app:anyK-sols}, we solve the recurrence relations and obtain all eigenstates.
In Sec.~\ref{app:anyK-Dq}, we calculate the multifractal dimensions and demonstrate that eigenstates display multifractal statistics.
Finally, in Sec.~\ref{asubsec:K2-approx}, we show the validity of the approximations used in calculating the multifractal dimensions.

\subsection{Choice of basis}
\label{app:anyK-basis}
\begin{figure*}[t]
\centering
\includegraphics[width=0.8\linewidth]{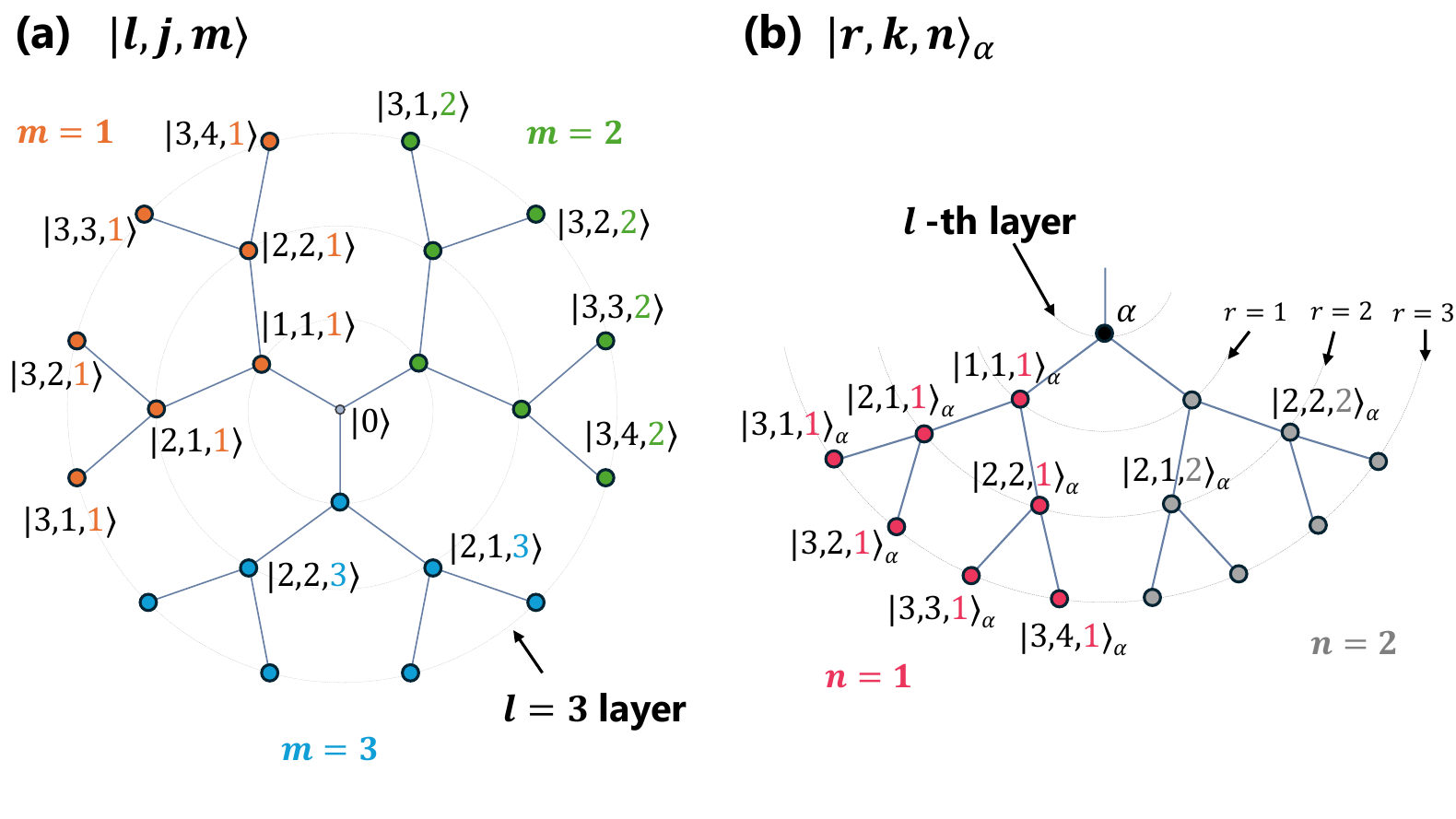} 
\caption{
Sketch of the basis construction in the case $K=2$.
(a) Symmetric basis. (b) Non-symmetric basis. 
The symmetric (non-symmetric) basis states are constructed by the position basis $\ket{l,j, m}$ ($\ket{r,k, n}_{\alpha}$).}
\label{fig:basis}
\end{figure*}
We consider the Cayley tree with connectivity $K$ and the total number of layers $M$.
Since the central node has $K+1$ branches and the $l$-th $(l=1, \cdots, M)$ layer of each branch has $K^{l-1}$ nodes, the dimension $\mathcal N$ of the Hilbert space is 
\begin{align}
    \mathcal{N} =  1 + (K+1) \times \sum_{l=1}^M K^{l-1} = 1 + (K+1) \frac{K^M -1}{K-1}.
\end{align}
By generalizing the discussion in Refs.~\cite{Aryal-IOP-2020,Mahan-2001-PRB}, we generate $\mathcal{N}$ orthonormal basis states which span the whole Hilbert space.

First, we choose the central node as one of the symmetric basis states:
\begin{align}
    |0) \coloneqq \ket{0}.
\end{align}
In this paper, $\ket{\cdots}$ denotes the position basis and $|\cdots)$ denotes the symmetric and non-symmetric basis states. 

Second, we generate the remaining symmetric basis states $| l)_m$ by symmetrizing the position basis $\ket{l,j,m}$ $(j=1, \cdots, K^{l-1})$ in the $l$-th $(l= 1,\cdots,M )$ layer of the branch $m$ $(m= 1,2,\ldots K+1)$ as
\begin{align}
\label{aeq: symm basis}
    | l)_m \coloneqq \frac{1}{\sqrt{K^{l-1}}} \sum_{j=1}^{K^{l-1}}  \ket{l,j,m},
\end{align}
which form $(K+1)M$ symmetric orthonormal states [see Fig.~\ref{fig:basis}~(a) for the case of $K=2$].

Third, we generate the remaining basis states, which we call non-symmetric basis states, as follows.
We choose a node $\alpha$ as the origin and consider the $K$ branches rooted at this node. 
When the node $\alpha$ is located in the $l$-th layer, the number of the remaining layers counting from $\alpha$ is $M-l$.
We specify the remaining layers as $r=1,2,\cdots,M-l$ [see Fig.~\ref{fig:basis}~(b)]. To define a non-symmetric basis state, we also need to choose a nontrivial $K$-th root of unity $\omega$.

We generate the non-symmetric basis $| l, r,\omega)_\alpha$ by weighting the position basis $\ket{r,k,n}_\alpha (k=1, \cdots,K^{r-1})$ in the $r$-th layer by the powers of $\omega$, where all states in a branch $n$ ($n=1, \cdots,K$) have the same weight:
\begin{align}
\label{aeq:nonsym}
    |l,r,\omega)_\alpha \coloneqq \frac{1}{\sqrt{K^{r}}} \sum^{K}_{n=1}\omega^n\sum^{K^{r-1}}_{k=1} \ket{r,k,n}_\alpha.
\end{align}
One can explicitly check that these states are orthonormal. The notation here is a little bit redundant since the node $\alpha$ determines the layer $l$. However, for convenience, we keep these indices in state description. 

An illustration of the construction of a non-symmetric basis state for
 $K=2$ is shown in Fig.~\ref{fig:basis}~(b). In the case of $K=2$, there is only one nontrivial root of unity $\omega=-1$, and as an example, we can write the states $|l,r=1,\omega=-1)$ and $|l,r=2,\omega=-1)$ explicitly:
\begin{align}
\label{aeq:nonsym-example1}
    |l,r=1,\omega=-1)_\alpha \coloneqq \frac{1}{\sqrt{2}} \Big(-
    \ket{1,1, 1}_\alpha + \ket{1,1, 2}_\alpha
    \Big),
\end{align}
\begin{align}
\label{aeq:nonsym-example2}
    |l,r=2,\omega=-1)_\alpha \coloneqq \frac{1}{\sqrt{2^2}} \Big(
    -\ket{2,1, 1}_\alpha - \ket{2,2, 1}_\alpha + \nonumber 
    \\+ \ket{2,1, 2}_\alpha +\ket{2,2, 2}_\alpha
    \Big).
\end{align}

We count the total number of non-symmetric basis states.
The origin $\alpha$ can be chosen from the $(K+1) \times K^{l-1}$ nodes of the $l$-th layer of the layer, and there are $K-1$ nontrivial roots of unity $\omega$.
Hence, we have 
\begin{align}
    \sum_{l=1}^{M-1} (K-1)(K+1) \times K^{l-1} (M-l) = \mathcal{N} -(K+1)M-1
\end{align}
non-symmetric basis states.
By combining the symmetric basis $|0)$, $|l)_m$ and the non-symmetric basis $|l,r,\omega)_\alpha$, we obtain $\mathcal{N}$ symmetry-adapted orthonormal basis states.
Therefore any state $\ket{\Psi}$ is expanded using these basis states as
\begin{align}
    \!\ket{\Psi} \!=\! \psi_0 |0) \!+\! \!\sum_{l=1}^M \!\sum_{m=1}^{M+1} \!\! \psi_{l,m} |{l})_m \!+\!\!\!
    \sum_{l=1}^{M{-}1} \!\sum_{r=1}^{M{-}l} \!\sum_{\omega} \!\sum_{\alpha \in \mathbb{G}_l} \!\!\phi_{l, r,\omega}^{\alpha} 
    |{l,\! r,\omega})_\alpha \!
\end{align}
where $\psi_0, \psi_{l,m}$ and $\phi_{l,r,\omega}^\alpha$ are the wave function components
and $ \mathbb{G}_l$ is 
the set of all $(K+1) \times K^{l-1}$ sites in the $l$-th layer. 
In this paper, we refer to eigenstates that can be expanded solely using symmetric (non-symmetric) basis states as symmetric (non-symmetric) eigenstates.

\subsection{Recurrence relations}
\label{app:anyK-recrel}
We employ the basis constructed in the previous section to find all solutions of the eigenvalue equation. We obtain two main cases of symmetric and non-symmetric eigenstates. The case of symmetric eigenstates, in turn, splits into two sub-cases when $\psi_0 =0$ and $\psi_0 \neq0$.

The eigenvalue equation $H \ket{\Psi} = E \ket{\Psi}$ 
is reduced to
\begin{align}
\begin{split}
\label{aeq:ev-sym}
    E \psi_{l,m} &= \sqrt{K} t_{\rm R} \psi_{l-1,m} +  \sqrt{K} t_{\rm L} \psi_{l+1,m} \\
    E \psi_0 &= t_{\rm L} \sum^{K+1}_{m=1} \psi_{1,m} \\ 
    0 &= \psi_{M+1,m} 
    \end{split}
\end{align}
with $\psi_{0,m} \coloneqq \psi_0/ \sqrt{K}$ for $l = 1, \ldots, M$ and $m=1,\ldots K+1$
in the symmetric basis and to
\begin{align}
\begin{split}
\label{aeq:ev-asym}
    E \phi_{l, r,\omega}^\alpha &= \sqrt{K} t_{\rm R} \phi_{l, r-1,\omega}^\alpha + \sqrt{K} t_{\rm L} \phi_{l, r+1,\omega}^\alpha \\ 
    0 &=  \phi_{l,0,\omega}^\alpha = \phi_{l,M-l+1,\omega}^\alpha 
    \end{split}
\end{align}
for $l = 1, \ldots, M-1$ and $r=1,2,\ldots,M-l$ in the non-symmetric basis. It is worth noting that the root of unity $\omega$ does not appear in the eigenvalue equations and is only responsible for an additional degeneracy.

\subsection{Solutions}
\label{app:anyK-sols}
We solve the recurrence relations~\eqref{aeq:ev-sym} and \eqref{aeq:ev-asym}.
\subsubsection{Symmetric eigenstates of Eq.~\eqref{aeq:ev-sym}}
\label{app-sub:K2 symmsol}
\begin{enumerate}
    \item \textit{The case of $\psi_0 = 0$} \\\\
    For $\psi_0 = 0$, $KM$ solutions are given in Sec.~\ref{subsec:K2-recur} in the main text.
    
    It should be noted that these solutions are $K$-fold degenerate. Nevertheless, the multifractal dimensions $D_q$ does not depend on a choice of a particular linear combination of the degenerate eigenstates, as we show in Appendix~\ref{app:anyK-Dq}.    
    \item \textit{The case of $\psi_0 \neq 0$}\\ \\
For $\psi_0 \neq 0$, Eq.~\eqref{aeq:ev-sym} has $M+1$ solutions.
By requiring that $\psi_{l,m}$ are equal for any $m$,
we rewrite Eq.~\eqref{aeq:ev-sym} as
\begin{align}
\begin{split}
\label{aeq:ev-sym-psi-neq0}
    E \psi_{l,m} &= \sqrt{K} t_{\rm L} \psi_{l-1,m} +  \sqrt{K} t_{\rm R} \psi_{l+1,m} \\
    E \psi_0 &= (K+1) t_{\rm L} \psi_{1,m}  \\
     0 &= \psi_{M+1,m}.
    \end{split}
\end{align}
for $l=1,2,\ldots, M$.

We use the ansatz
\begin{align}
\begin{split}
\label{aeq:ansatz-antisym}
    E &= 2 \sqrt{Kt_{\rm L} t_{\rm R}} \cos{\theta},\\ 
    {\psi_0} &=  \sin{\delta},  \\ 
    \psi_{l,m} &=  \beta^l \sin{(l \theta + \gamma)} 
      \end{split}
\end{align}
with unknown parameters 
$\theta, \delta, \gamma$, and $\beta=\sqrt{t_{\rm R} /t_{\rm L}}$. The phase for $\psi_0$ is different from $\gamma$, because $\psi_0$ appears in two different equations:
\begin{align}
\begin{split}
E\psi_0=(K+1)t_{\rm L} \psi_{1,m}, \\
E\psi_{1,m}=t_{\rm L} \psi_{0}+\sqrt{K} t_{\rm R} \psi_{2,m}.
\end{split}
\end{align}
These equations give
\begin{align}
\begin{split}
   2\sqrt{K}\cos(\theta)\sin(\delta)=(K+1)\sin(\theta+\gamma) ,\\  2\sqrt{K}\cos(\theta)\sin(\theta+\gamma)=\beta^{-2}\sin(\delta)+\beta^2 \sqrt{K}\sin(2\theta+\gamma).
   \end{split}
\end{align}
We can find the expression for $\sin{(\delta)}$ from the second line. Substituting it into the first line, we have
\begin{align}
    2\beta^2 K \cos(\theta)[2\cos(\theta)\sin(\theta+\gamma)-\beta^2 \sin(2\theta+\gamma)] \nonumber \\
    =(K+1)\sin(\theta+\gamma),
\end{align}
leading to
\begin{align}
    2\beta^2 (1-\beta^2) K [\sin(2\theta)+\cos(2\theta)\tan(\gamma)]+2\beta^2 K \tan(\gamma) \nonumber \\
    =(K+1)[\tan(\theta)+\tan(\gamma)].
\end{align}
Finally, we obtain
\begin{align}
\label{aeq:gamma-omega1}
    &\tan{\gamma}=  \nonumber \\
    & \frac{[(K+1)(1+\tan^2 (\theta))-4 \beta^2 K (1-\beta^2)]\tan(\theta)}{[2\beta^2\!K-\!(K+1)](1+\tan^2(\theta))+2 \beta^2 K (1-\beta^2 )(1-\tan^2(\theta))}.
\end{align}
We note that for Hermitian limit $\beta=1$, we reproduce the results of Ref.~\cite{Mahan-2001-PRB}.

From the third line 
of Eq.~\eqref{aeq:ev-sym-psi-neq0}, we have
\begin{align}
\label{aeq:gamma-omega2}
    \sin{[ (M+1) \theta + \gamma ]} = 0.
\end{align}
Combining Eqs.~\eqref{aeq:gamma-omega1} and ~\eqref{aeq:gamma-omega2}, we obtain the $M+1$ solutions for $\theta_n$ and $\gamma_n ~(0<\theta_n < \pi, 0<\gamma_n < \pi, n=1,\cdots,M+1)$.
Notably, $\psi_{l,m}$ in Eq.~\eqref{aeq:ansatz-antisym} follows a similar 
dependence on $\beta$ (namely $\psi_{l,m} \propto \beta^l $) as found for symmetric eigenstates with $\psi_0=0$, given by Eq.~\eqref{eq:sol-psi0}.

As in the previous case, we discuss the degeneracy of the obtained solutions. 
Since $\theta_n / \pi$ (except one solution $\theta=\pi/2$) should be irrational numbers, the $M$ eigenstates are nondegenerate. 
Furthermore, in this case, there is also no degeneracy, neither with other symmetric or non-symmetric states. 
Hence multifractal dimensions of these $M$ solutions are well-defined.

\end{enumerate}

\subsubsection{Non-symmetric eigenstates of Eq.~\eqref{aeq:ev-asym}}
Equation~\eqref{aeq:ev-asym} is nothing but the eigenvalue equation of the Hatano-Nelson model~\cite{Hatano-PRL-1996, *Hatano-PRB-1997} with system size $M-l$ under open boundary conditions, whose $M-l$ solutions are given by (see Appendix~\ref{app:hn})
\begin{align}
\label{aeq:sol-asym}
    \phi_{l,r, \omega}^{\alpha \, (n)} = \beta^r \sin{(\theta_{n,l} r)}, \quad E_{n,l} = 2 \sqrt{K t_{\rm R} t_{\rm L}} \cos{\theta_{n,l}}
\end{align}
with 
\begin{align}\label{aeq:anti-theta}
     \quad \theta_{n,l} \!=\! \frac{n \pi}{M{-}l{+}1} ~(n\!=\!1,2,\ldots,M{-}l;\, l\!=\!1,2,\ldots,M).
\end{align}
Here, $E_{n,l}$ does not depend on the choice of $\alpha \in { \mathbb{G}_l}$, and the root of unity $\omega$.
Since $\mathbb{G}_l$ is 
the set of all sites in the $l$-th layer, it has $(K+1) \times K^{l-1}$ elements, and there are $K-1$ different nontrivial roots of unity, the eigenstates are $(K+1)(K-1) \times K^{l-1}$-fold degenerate.
Therefore we have 
\begin{align}
    \sum_{l=1}^M (M-l) \times (K+1) \times K^{l-1} = \mathcal{N} -1 -(K+1)M
\end{align}
solutions of Eq.~\eqref{aeq:ev-asym}.

By combining the solutions in Eqs.~\eqref{aeq:ev-sym}~and~\eqref{aeq:ev-asym},
we have
\begin{align}
    (KM) + (M + 1) + (\mathcal{N} - 1 -(K+1)M) = \mathcal{N}
\end{align}
linearly independent solutions, which coincides with the dimension of the Hilbert space $\mathcal{N}$.

Again we discuss the degeneracy of obtained solutions.
The degeneracy of the non-symmetric eigenstates is more complicated than for the symmetric eigenstates. It is also more important since this degeneracy can alter the multifractal dimensions $D_q$.

First, we discuss the degeneracy between non-symmetric eigenstates. 
As described below in Eq.~\eqref{aeq:anti-theta}, the eigenstates are $(K+1)(K-1) \times K^{l-1}$-fold degenerate with others whose basis origin $\alpha$ are in the same layer.
Additionally, they can be degenerate with  eigenstates whose basis origin $\alpha$ are in a different layer, i.e., we have sometimes $E_{n,l} = E_{n',l'}$ for different pairs $(n,l)$ and $(n',l')$ equivalent to 
\begin{align}
\label{aeq:sol-asym-pair}
  \frac{n }{M-l +1} = \frac{n' }{M-l' +1}.
\end{align}
This degeneracy allows for a linear combination of eigenstates whose layers of the basis origin are different. 
This kind of degeneracy can alter multifractal dimensions, and thus we cannot determine the values $D_q$ of them uniquely. 
Indeed, we explicitly demonstrate that the multifractal dimensions depend on the choice of linear combinations in the main text [Fig.~\ref{fig:disorder} (a, b)].

Second, we note that the non-symmetric eigenstates can also be degenerate with symmetric eigenstates. In this case, $M-l+1$ and $M+1$ should have common divisors. However, this is prohibited if we choose $M+1$ as a prime number. 

If there is no degeneracy between non-symmetric and symmetric states, the $(K+1)M$ symmetric eigenstates have 
well-defined multifractal dimensions described in the main text.

\subsection{Calculating the multifractal dimensions $D_q$}
\label{app:anyK-Dq}
We calculate the multifractal dimensions $D_q$ of the eigenstates obtained above. 
\subsubsection{\texorpdfstring{$D_q$}{Dq} for symmetric eigenstates of Eq.~\eqref{aeq:ev-sym}}

\noindent 1.~~\textit{The case of $\psi_0 = 0$} \\\\
In the general solution, given by Eq.~\eqref{eq:sol-sym} in the main text, we first approximate the symmetric eigenstates as plane waves (see Appendix~\ref{asubsec:K2-approx}).
This simplification leads to 
    \begin{align}
    \label{aeq:I-wv-approx-K2-sym1}
        \psi_{l,m}^{(n)} \simeq \frac{1}{\sqrt{C}}(\beta e^{\ii \theta_n})^l
    \end{align}
    with the normalization constant
\begin{align}
    C =  \sum_{l=1}^M \beta^{2l} = \frac{ \beta^2 (\beta^{2M}-1)}{\beta^2 -1}.
\end{align}
To calculate $D_q$, we go back to the position basis $\ket{l,j,m}$.
For instance, the solution in Eq.~\eqref{eq:sol-sym} can be expanded in the position basis as
\begin{align}
\ket{\Psi_n} &\simeq
    \sum_{l=1}^M 
    \frac{ (\beta e^{\ii \theta_n})^l }{\sqrt{K^{l-1} C}} \times 
    \nonumber \\
    \Bigg[&c_1  
    \Big(
    \ket{l,1,1} + \cdots + \ket{l,K^{l-1},1}
    \Big) \nonumber \\
    + &c_2 \Big(
    \ket{l,1,2} + \cdots + \ket{l,K^{l-1},2}
    \Big)+\ldots \nonumber \\
     + &c_{K+1} \Big(
    \ket{l,1,K+1} + \cdots + \ket{l,{K^{l-1}},K+1}
    \Big)
    \Bigg].
\end{align}
Without loss of generality, we have assumed $\sum_{i=1}^{K+1} \abs{c_i}^{2}=1$. 
Then the inverse participation ratio $I_q =\sum_{l,j,m} \abs{\bra*{l,j,m}\ket*{\Psi_n} }^{2q} $ in the position basis becomes
\begin{align}\label{aeq:Iq-cont}
    I_q = \left( \sum_{i=1}^{K+1} \abs{c_i}^{2q} \right)  \frac{K^{q-1}}{C^q} \sum_{l=1}^M \left(\frac{\beta^{2q}}{K^{q-1}}\right)^l 
\end{align}
leading to 
\begin{align}
\label{aeq:Iq-sym}
    I_q=  \left[
    \frac{\beta^2 -1}{ \beta^2 (\beta^{2M}-1)}
    \right]^q
    \frac{
    1-\left(\frac{\beta^{2q}}{K^{q-1}}\right)^M 
    }{\beta^{-2q}-{K^{1-q}}} \sum_{i=1}^{K+1} \abs{c_i}^{2q}.
\end{align}

For the sake of simplicity in notation, we introduce the dimensionless parameter $q^*$ as
\begin{align}
    \label{aeq: def of q*}
    q^* \coloneqq \frac{\log{K}}{\log{K}-\log{(\beta^2)}}.
\end{align}
Note in passing that Eq.~\eqref{aeq:Iq-cont} yields the logarithmic correction $(\propto \ln{M})$ for the participation entropy at $q=q^*$~\cite{Luitz-14}.
Depending on the value $\beta$, we have the following four cases.
\begin{enumerate}
    \item[(1)] For $\beta>1$ and $\beta^{2q}/ K^{q-1} < 1$, which are equivalent~to 
    \begin{align}
    &{\rm i)}~ 1<\beta < \sqrt{K} {~\rm and~} q > q^*, \ {\rm or} \nonumber \\
    &{\rm ii)}~ \sqrt{K} <\beta  {~\rm and~} q < q^*,  \nonumber 
     \end{align}
    the inverse participation ratio $I_q$ becomes
    \begin{align}
        I_q 
        \rightarrow \mathcal{N}^{-q \frac{\log{(\beta^2)}}{\log{K}}}
    \end{align}
    in the limit $\mathcal{N} \rightarrow \infty$. Hence, we obtain $\tau_q = (q \log{\beta^2})/{\log{K}}$. \\
    \item[(2)] For $\beta>1$ and $\beta^{2q}/{K^{q-1}} > 1$, which are equivalent~to 
    \begin{align}
        &{\rm i)}~ 1<\beta < \sqrt{K} {~\rm and~} q < q^*, \ {\rm or} \nonumber \\
         &{\rm ii)}~ \sqrt{K} <\beta  {~\rm and~} q > q^*, \nonumber
    \end{align}
    the inverse participation ratio $I_q$ becomes 
    \begin{align}
         I_q \rightarrow \mathcal{N}^{-(q-1)} 
    \end{align}
    in the limit $\mathcal{N} \rightarrow \infty$. Hence, we obtain $\tau_q = q-1$. \\
    \item[(3)] For $ \beta<1$ and $\beta^{2q}/ K^{q-1} < 1$, which are  equivalent~to 
    \begin{align}
        0 < \beta < 1 {~\rm and~} q > q^*, \nonumber
    \end{align} 
    the inverse participation ratio $I_q$ becomes 
     \begin{align}
     I_q \rightarrow \mathcal{N}^{0}
    \end{align}
    in the limit $\mathcal{N} \rightarrow \infty$. Hence, we obtain $\tau_q = 0$. \\
    \item[(4)] For $\beta<1$ and $\beta^{2q}/ K^{q-1} > 1$, which are equivalent~to
    \begin{align}
        0 < \beta < 1 {~\rm and~} q < q^*, \nonumber
    \end{align}
    the inverse participation ratio $I_q$ becomes 
    \begin{align}
      I_q \rightarrow \mathcal{N}^{-\frac{q-q^*}{q^*}}
    \end{align}
    in the limit $\mathcal{N} \rightarrow \infty$. Hence, we obtain $\tau_q = (q/q^*) -1$.
\end{enumerate}
We note that the multifractal dimensions do not depend on the choice of $c_i$ ($i=1,\cdots,K+1$).
Rearranging the above results, the multifractal dimensions $D_q = \tau_q / (q-1)$ are obtained as follows:
\begin{enumerate}
   \item[(1)] For $0 < \beta < 1$ \\
   \begin{equation} 
   \label{aeq:Dq01}
       D_q = 
       \begin{cases}
           1 - \frac{q}{q-1} \frac{\log (\beta^2)}{\log K} \quad &(q <q^*) \\
           0 \quad &(q > q^*).
       \end{cases}
   \end{equation}
   \item[(2)] For $1 < \beta < \sqrt{K}$ \\
   \begin{equation}
    \label{aeq:Dq12}
       D_q = 
       \begin{cases}
           1 \quad &(q <q^*) \\
          \frac{q}{q-1} \frac{\log (\beta^2)}{\log K} \quad &(q > q^*).
       \end{cases}
   \end{equation}
   \item[(3)] For $\sqrt{K} < \beta$ \\
   \begin{equation}
    \label{aeq:Dq2-}
       D_q = 1.
   \end{equation}
\end{enumerate}
The above results are discussed in the main text [see Eqs.~\eqref{eq:Dq1},~\eqref{eq:Dq2} and \eqref{eq:Dq3}]. \\

\noindent 2.~~\textit{The case of $\psi_0 \neq 0$} \\\\
In Eq.~\eqref{aeq:ansatz-antisym}, we again approximate (see Appendix~\ref{asubsec:K2-approx})
\begin{align}
 \label{aeq:I-wv-approx2}
    \psi_{l,m}^{(n)} \simeq  \frac{1}{\sqrt{C}}(\beta e^{\ii \theta_n})^l
\end{align}
with $C = \beta^2(\beta^{2M}-1)/(\beta^2 -1)$, similarly to Eq.~\eqref{aeq:I-wv-approx-K2-sym1}.
Therefore, the multifractal dimensions $D_q$ take the same value as those in the case of $\psi_0 = 0$ in Eqs.~(\ref{aeq:Dq01})--(\ref{aeq:Dq2-}).

For multifractal dimensions to be uniquely determined, the considered states should not degenerate with non-symmetric solutions. This degeneracy can lead to the possibility that multifractal dimensions may depend on the linear combination of the degenerate states.
We can avoid that issue by choosing $M+1$ as a prime number. In this case, the multifractal dimensions of the $(K+1)M$ symmetric solutions can be uniquely determined, independent of the linear combination of the degenerate states.

\subsubsection{$D_q$ for non-symmetric eigenstates of Eq.~\eqref{aeq:ev-asym}}
The non-symmetric eigenstates are highly degenerate, and their multifractal dimensions depend on the choice of the linear combination of degenerate states.
Here, we show the multifractal dimensions of the particular eigenstates in Eq.~(\ref{aeq:sol-asym}) take the same value as symmetric eigenstates by choosing the specific linear combinations.
We focus on solutions where the layer of the origin $\alpha$ satisfies $\lim_{M \rightarrow \infty} (l/M) < 1$, meaning that the origin $\alpha$ is far from the boundary.
Since these modes spread over the tree in the Hermitian limit ($\beta = 1$), they should be related to the skin effect.
Similarly to Eq.~\eqref{aeq:I-wv-approx-K2-sym1}, we first approximate (see Appendix~\ref{asubsec:K2-approx})
    \begin{align}
    \label{aeq:III-wv-approx3}
        {\phi_{l,r,\omega}^{\alpha}}^{(n)} \simeq \frac{1}{\sqrt{C}}(\beta e^{\ii \theta_n})^r
    \end{align}
    with the normalization constant
\begin{align}
    C =  \sum_{r=1}^{M-l} \beta^{2r} = \frac{ \beta^2 (\beta^{2(M-l)}-1)}{\beta^2 -1}.
\end{align}
To calculate $D_q$, we go back to the position basis $\ket{r,k,n}_{\alpha}$.
The solution 
\begin{align}
    \ket{\Psi_n} =\sum_{r=1}^{M-l} {\phi_{l,r,\omega}^\alpha}^{(n)} |l,r,\omega)_\alpha
\end{align}
can be expanded on the position basis $\ket{r,k, n}_{\alpha}$ as
\begin{align}
\ket{\Psi_n} &= 
\sum_{r=1}^{M-l} 
\frac{\phi_{l,r,\omega}^\alpha}{\sqrt{K^{r}}}
\sum^{K}_{n=1}\omega^n \Big(
    \ket{r,1,n} + \cdots + \ket{r,K^{r-1},n}
    \Big)
\end{align}
Hence, the inverse participation ratio $I_q$ in the position basis becomes
\begin{align}
    I_q
    &= \sum_{r,k,n} \abs{\bra*{r,k, n}\ket*{\Psi_n}}^{2q} \nonumber \\
    &= \frac{1}{C^q} \sum_{r=1}^{M-l} \left( \frac{\beta^{2q}}{K^{q-1}}\right)^r \nonumber  \\
    &= \left[
    \frac{\beta^2 -1}{ \beta^2 (\beta^{2(M-l)}-1)}
    \right]^q
    \frac{{K^{1-q}}
    \left[1-\left(\frac{\beta^{2q}}{K^{q-1}}\right)^{M-l} \right]
    }{\beta^{-2q}-{K^{1-q}}}. \label{aeq:IPR-III}
\end{align}
Identifying $M-l$ with $M$ essentially reduces Eq.~\eqref{aeq:IPR-III} to Eq.~\eqref{eq:Iq-sym}.
Therefore, when the value $l$ satisfies $\lim_{M \rightarrow \infty} (l/M) < 1$,
the multifractal dimensions $D_q$ take the same value as in the case of symmetric eigenstates in the thermodynamic limit $M \rightarrow \infty$.

As mentioned above, the multifractal dimensions are not well-defined for non-symmetric eigenstates because they depend on the choice of linear combinations of the degenerate eigenstates.
Nevertheless, we have shown that the specific linear combinations such that the eigenstate is monotonically distributed from the origin $\alpha$ yield the same multifractal dimensions as symmetric eigenstates.

\subsection{Approximation of eigenstates}
\label{asubsec:K2-approx}

Here, we show the validity of the approximations used in Eqs.~\eqref{aeq:I-wv-approx-K2-sym1},~\eqref{aeq:I-wv-approx2} and \eqref{aeq:III-wv-approx3}.
For simplicity, we consider a state given by 
\begin{align}
\label{aeq:approx-es}
    \ket{\Psi} = \frac{1}{\sqrt{C}} \sum_{l=1}^M \frac{\beta^l}{\sqrt{K^l}}  \sin{(\theta l)} 
     \Bigg[
    \ket{l,1} + \cdots + \ket{l,K^{l}}
    \Bigg] 
\end{align}
with the normalization constant $C = \sum_{l=1}^M \sin^2 {(\theta l)} \beta^{2l} $.
Multifractal dimensions
\begin{align}
    D_q = \lim_{\mathcal{N} \to \infty} \frac{1}{1-q} \frac{1}{\log{\mathcal{N}}} \log{I_q}
\end{align}
are determined by the asymptotic behavior of the inverse participation ratio $I_q$:
\begin{align}
\label{aeq:approx-IPR}
    I_q =  \sum_{l=1}^M \sum_{j=1}^{K^{l}} \abs{ \bra*{{l,j}}\ket*{\Psi}}^{2q}
\end{align}
in the limit of large $\mathcal{N}$.
By substituting Eq.~\eqref{aeq:approx-es} into Eq.~\eqref{aeq:approx-IPR}, $\log I_q$ becomes
\begin{align}
\label{eq:Iq_with_phases}
\log I_q = \log\Big[ \sum^{M}_{l=1} \sin^{2q}(\theta l)\Big(\frac{\beta}{\sqrt{K}}\Big)^{2ql}K^l\Big] \nonumber \\
-q\log\Big[\sum^{M}_{l=1} \sin^{2}(\theta l)\beta^{2l}\Big]
\end{align}
where the second term arises from the normalization constant.

Let us consider the case where
$q>q^*$ and $\beta>1$ for simplicity [$q^*=\log{K}/(\log{K}-\log{\beta^2)}$]. 
In this case, the exponential multipliers $(\beta/\sqrt{K})^{2q}K$ and $\beta$ in both sums in Eq.~\eqref{eq:Iq_with_phases} are greater than $1$.
The terms under the sums are the product of a bounded function and the exponentially growing functions, and therefore the main asymptotics of $\log I_q$ is given by the largest term in the sums since multiplication by a bounded function does not change the exponential character of the asymptotics:
\begin{align}
\log I_q\sim \log \Big[\sin^{2q}(\theta M) \Big(\frac{\beta}{\sqrt{K}}\Big)^{2qM}K^M\Big] \nonumber \\
-q\log \Big[\sin^2 (\theta M)\beta^{2M} \Big].
\end{align}
After moving $\sin^{2q}(\theta M)$ out of the logarithms, we find that the terms are canceled out, and hence the inverse participation ratio $I_q$ only depends on the exponential factors: 
\begin{align}
\log I_q\sim \log \Big[ \Big(\frac{\beta}{\sqrt{K}}\Big)^{2qM}K^M\Big]-q\log \Big[\beta^{2M}\Big].
\end{align}
Therefore, the sine term is irrelevant to the multifractal dimensions.
When an exponential multiplier is less than $1$, the corresponding sum is convergent and does not contribute to the multifractal dimensions.
From a physical perspective, the irrelevance of the sine term indicates that it does not change the localization properties.

\section{Lack of multifractal statistics in conventional single-particle skin effects}\label{app:higher-order}
We demonstrate the absence of multifractality in conventional single-particle skin effects by analyzing a simple model.
Consider a $d$-dimensional hypercubic lattice with side length $L$ ($\mathcal{N} = L^d$).
Let us examine a state that occupies an 
$n$-dimensional region (e.g., $(d-n)$-th order non-Hermitian skin mode~\cite{Kawabata-PRB-20,Okugawa-20}).
As a typical example, we consider the state described by
\begin{align}
    \psi(x_1, \cdots,x_d) = \left(
    \prod\limits_{j=1}^{d-n}
    \frac{(\beta_j)^{x_j}}{\sqrt{C(\beta_j)}}
    \right)
    \left(
    \prod\limits_{j=d-n+1}^{d}
    \frac{1}{\sqrt{L}}
    \right)
\end{align}
with the normalization constant
\begin{align}
    C(\beta) =  \sum_{j=1}^L \beta^{2j} = \frac{ \beta^2 (\beta^{2L}-1)}{\beta^2 -1}.
\end{align}
This state is exponentially localized in the direction of $x_j$ ($j=1,\cdots,d-n$), but extends in the remaining directions.
This wave function is essentially the same as that assumed in the non-Bloch band theory including higher dimensions~\cite{Amoeba-24}, and thus it is expected to capture the characteristics of the skin effect accurately.
For simplicity, we assume $\beta_j>1$ in the following discussion.
The inverse participation ratio $I_q$:
\begin{align}
    I_q = \sum_{x_1, \cdots,x_d = 1}^L |\psi(x_1, \cdots,x_d)|^{2q}
\end{align}
is calculated as 
\begin{align}
    I_q = 
    \left(
    \prod\limits_{j=1}^{d-n}
    \sum_{x_j = 1}^L 
    \frac{(\beta_j)^{2q x_j}}{{C(\beta_j)}^q}
    \right)
    \left(
    \prod\limits_{j=d-n+1}^{d}
     \sum_{x_j = 1}^L
    \frac{1}{{L}^q}
    \right).
\end{align}
From a straightforward calculation, we have 
\begin{align}
    \sum_{x = 1}^L 
    \frac{(\beta_j)^{2q x}}{{C(\beta)}^q} 
    &=\left[
    \frac{\beta^2 -1}{\beta^2 (\beta^{2L}-1)}
    \right]^q
    \frac{\beta^{2q}(\beta^{2qL}-1)}{\beta^{2q}-1}.
\end{align}
The factor 
\begin{align}
     \frac{\beta^{2qL}-1}{(\beta^{2L}-1)^q}
\end{align}
is independent of side length $L$ for $L \rightarrow \infty$.
Therefore the multifractal dimension is calculated as
\begin{align}
    D_q = \lim_{\mathcal{N} \to \infty} \frac{1}{1-q} \frac{1}{\log \mathcal{N}} \log L^{(1-q)n} = \frac{n}{d}.
\end{align}
While the fractality $0<D_q <1$ can appear, this state can be captured by a single exponent, and therefore does not exhibit multifractal statistics.
This contrasts with the skin modes on a tree, for which the multifractal dimensions depend on the choice of $q$.

\section{Effect of strong disorder} \label{app:strong}
\begin{figure}[h]
\centering
\includegraphics[width=0.75\linewidth]{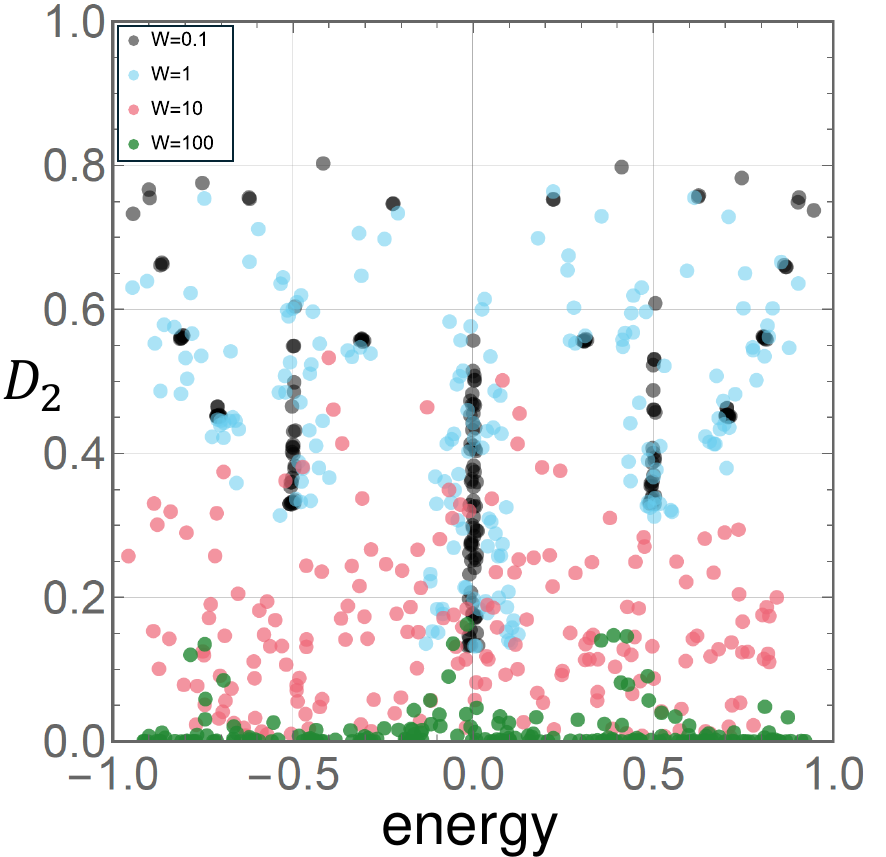} 
\caption{The multifractal dimension $D_2$ as a function of the rescaled energy for a Cayley tree in the presence of disorder with $K=2$, $M+1=7$, $t_{\rm R}=1.1$ and $t_{\rm L}=0.9$ ($\beta=\sqrt{11/9}$).
The disorder strength is chosen $W=0.1$ (black), $W=1$ (blue), $W=10$ (red), $W=100$ (green).
To facilitate data comparison, every energy is rescaled by $E/1.05\sqrt{8 t_{\rm R}t_{\rm L}}$ for $W=1$,
$E/2.25\sqrt{8 t_{\rm R}t_{\rm L}}$ for $W=10$, and $E/18\sqrt{8 t_{\rm R}t_{\rm L}}$ for $W=100$ respectively.}
\label{afig:strong}
\end{figure}
We provide the additional numerical results for the disordered Hamiltonian in the main text:
\begin{align}
    H_\textrm{dis}=H + \sum_{j=1}^\mathcal{N} U_j \ket{j}\bra{j},
\end{align}
where $H$ is the nonreciprocal Hamiltonian in Eq.~\eqref{eq:hamiltonian} and $U_j$ takes a random value for each site ($U_j \in [-u, u]$, $u=W\Delta$ with $W \geq 0$ and the energy interval $\Delta = \sqrt{4K t_{\rm R}t_{\rm L}}/M$).
Figure~\ref{afig:strong} shows the multifractal dimension $D_2$ of each eigenstate for different disorder strength $W$.
When the disorder strength is increased from $W=0.1$ to $W=1$,
the number of eigenstates whose multifractal dimension is around $D_2 \in [0.7,0.8]$ decreases.
Nevertheless, most of the multifractal dimension $D_2$ still take $D_2 \in [0.15,0.8]$ indicating that the multifractality $ 0 < D_2 < 1$ persists for $W=1$.
For $W = 10$, the value of $D_2$ significantly decreases for almost all eigenstates. 
This behavior of $D_2$ suggests a competition between non-Hermiticity and disorder leads to intricate occupation of skin modes on the tree, which merits further study.
For $W = 100$, the strong disorder destroys the multifractal properties of eigenstates ($D_2 = 0$), which implies the eigenstates are strongly localized on specific branches.

\bibliography{ref.bib}

\end{document}